\documentclass[10pt,journal]{IEEEtran}
    \usepackage{enumerate}
    \usepackage{subfigure}
	\usepackage{amssymb}
	\usepackage{graphicx}
	\usepackage{epstopdf}
	\usepackage{gensymb}
	\usepackage{color}
	\usepackage{amsmath}
	\usepackage{enumerate}
	\usepackage{cite}
	\usepackage{comment}
	\usepackage{amsthm}
	\usepackage{enumitem}
	\theoremstyle{definition}
	\usepackage{soul}
	
	\usepackage[framemethod=TikZ]{mdframed}
	
	\usepackage{epstopdf}
	
	\usepackage{algorithmicx}
	\usepackage[ruled]{algorithm}
	\usepackage{algpseudocode}
	\alglanguage{pseudocode}
    
	\usepackage{array}
	\usepackage{booktabs}
	\usepackage{multirow}

	\usepackage{threeparttable} 
	\usepackage{hyperref}
	
	\usepackage{mathtools}

	\usepackage{url}
	\urldef{\mailsa}\path|{alfred.hofmann, ursula.barth, ingrid.haas, frank.holzwarth,|
		\urldef{\mailsb}\path|anna.kramer, leonie.kunz, christine.reiss, nicole.sator,|
		\urldef{\mailsc}\path|erika.siebert-cole, peter.strasser, lncs}@springer.com|

\definecolor{garrisonpink1}{rgb}{0.858, 0.188, 0.478}

\newcommand{\mypara}[1]{\vspace{2pt}\noindent\textbf{{#1: }}}
\definecolor{PangLHpink2}{rgb}{0.2, 0.2, 1}

\definecolor{yjr}{RGB}{100,149,237}

\definecolor{myblue}{rgb}{0,0,0}

\begin{document}

\title{MLMSA: \textbf{M}ulti-\textbf{L}abel \textbf{M}ulti-\textbf{S}ide-Channel-Information enabled Deep Learning \textbf{A}ttacks on APUF Variants}

\author{
Yansong Gao ({\it Member IEEE}), Jianrong Yao, Lihui Pang\IEEEauthorrefmark{2}, Wei Yang, Anmin Fu, \\ Said F.~Al-Sarawi ({\it Senior Member IEEE}), and Derek Abbott ({\it Fellow IEEE})

\IEEEcompsocitemizethanks{\IEEEcompsocthanksitem Y.~Gao, J.~Yao, W.~Yang, and A.~Fu are with School of Computer Science and Engineering, Nanjing University of Science and Technology, Nanjing, China. 
Y.~Gao is also with Data61, CSIRO, Australia. e-mail: 
\{120106222744,yansong.gao;generalyzy;fuam\}@njust.edu.cn.}

\IEEEcompsocitemizethanks{\IEEEcompsocthanksitem L. Pang\IEEEauthorrefmark{2} (corresponding author) is with School of Electrical Engineering, University of South China, Hengyang, China, and security engineering laboratory, Sungkyunkwan University, Korea. e-mail: sunshine.plh@hotmail.com}

\IEEEcompsocitemizethanks{\IEEEcompsocthanksitem  S.~Al-Sarawi and D.~Abbott are with School of Electrical and Electronic Engineering, The University of Adelaide, Adelaide, SA 5005, Australia. e-mail: \{said.alsarawi;derek.abbott\}@adelaide.edu.au}

}

\maketitle

\begin{abstract}
To improve the modeling resilience of silicon strong physical unclonable functions (PUFs), in particular, the APUFs that yield a very large number of challenge response pairs (CRPs), a number of composited APUF variants such as XOR-APUF, interpose-PUF (iPUF), feed-forward APUF (FF-APUF), and OAX-APUF have been devised. When examining their security in terms of modeling resilience, utilizing multiple information sources such as power side channel information (SCI) or/and reliability SCI given a challenge is under-explored, which poses a challenge to their supposed modeling resilience in practice. Building upon multi-label/head deep learning model architecture, this work proposes Multi-Label Multi-Side-channel-information enabled deep learning Attacks (MLMSA) to thoroughly evaluate the modeling resilience of aforementioned APUF variants. Despite its simplicity, MLMSA can successfully break large-scaled APUF variants, which has not previously been achieved. More precisely, the MLMSA breaks 128-stage $30$-XOR-APUF, ($9,9$)- and ($2,18$)-iPUFs, and ($2,2,30$)-OAX-APUF when CRPs, power SCI and reliability SCI are concurrently used. It breaks 128-stage $12$-XOR-APUF and ($2,2,9$)-OAX-APUF even when only the easy-to-obtain reliability SCI and CRPs are exploited. The 128-stage six-loop FF-APUF and one-loop 20-XOR-FF-APUF can be broken by simultaneously using reliability SCI and CRPs. All these attacks are normally completed within an hour with a standard personal computer. Therefore, MLMSA is a useful technique for evaluating other existing or any emerging strong PUF designs.

\end{abstract}

\IEEEpeerreviewmaketitle


\section{Introduction}\label{Sec:Intro}

Physical unclonable functions (PUFs) provide hardware instance-specific outputs (known as responses) to queried inputs (known as challenges), thus challenge-response-pairs (CRPs) generally function as `fingerprints' of hardware devices~\cite{herder2014physical,gao2020physical,liu2019xor}. PUFs can be categorized based on the number of yielded CRPs into weak and strong PUFs~\cite{herder2014physical,gao2020physical}. Weak PUFs have a limited number of CRPs which must be protected, so that its primary application is volatile key provision~\cite{gao2018lightweight,gao2021noisfre}. On the other hand, strong PUFs offer a very large number of CRPs, which can be used in many security applications ranging from identification, lightweight authentications to oblivious transfer~\cite{gao2020physical}. Among strong PUFs, the arbiter PUF (APUF)~\cite{gassend2002silicon,suh2007physical,zalivaka2018reliable} is the most studied design due to its compactness and compatibility with silicon fabrication processes. 
However, the APUF is vulnerable to modeling attacks due to its linear structure. To increase the complexity of modeling attacks, various non-linearity injection techniques have been used to construct APUF variants including the representative $l$-XOR-APUF, ($x,y$)-Iterpose PUF (iPUF)~\cite{nguyen2019interpose}, feed-forward APUF (FF-APUF)~\cite{lim2005extracting}, and ($x,y,z$)-OAX-APUF~\cite{yao2022design}. 
These APUF variants are resilient to modeling attacks to a large extent given that their scale is increased (i.e., 128-stage or/and large number of underlying APUFs composited)~\cite{gao2022systematically} when accessing to high performance computing platform, e.g., server with a cluster of GPUs and resourceful memory is unavailable.

\noindent{\bf State-of-the-Art of Modeling Attacks:} Majority of modeling attacks exploit CRPs \textit{only} to train the model. By using deep learning (DL) (i.e., multiple layer perception network), purely CRP based modeling attacks can break $128$-stage $7$-XOR-APUF, $64$-stage ($11,11$)-iPUF, $128$-stage FF-APUF with $5$ loops, $64$-stage $6$-XOR-FF-APUF with $5$ loops~\cite{wisiol2021neural}. In addition, it has been shown that the side-channel information (SCI) including unreliability~\cite{delvaux2013side,becker2015gap}, power or timing~\cite{ruhrmair2014efficient}, and photonic emission~\cite{tajik2017photonic} can be utilized to model the APUF or its variants. However, merely relying on SCI is insufficient to break APUF variants once it is properly scaled, and acquisition of some SCIs, e.g., photonic emission and timing require costly peripheral equipment.

To date, little efforts have been paid to hybrid modeling attacks on strong PUFs, in particular, APUF variants. The two hybrid attacks that are multi-class/single-label multi-SCI attack (SLMSA)~\cite{liu2022multiclass}, and gradient-based reliability hybrid attack (GRA)~\cite{tobisch2021combining} use not only CRP but also SCIs concurrently to break the APUF variants at a larger scale. 
However, there are still limitations in these attempts. More specifically, the SLMSA has a large dimension as its trained model output dimension is a multiplication per SCIs (SCI including the binary response information). 
In addition, this work mainly examines the efficiency of CRP as well as power SCI hybrid attacks, but efficacy when easy-to-obtain unreliability SCI is available has not been explored by~\cite{liu2022multiclass}. The reason is that Liu \textit{et al.}~\cite{liu2022multiclass} recognized the dimension of the reliability SCI could be much higher, potentially resulting in a dimensional curse (detailed in Section~\ref{sec:curse}). For the GRA specifically devised to attack iPUFs, it requires the differential mathematical model of underling APUFs, which is non-trivial to adopt without in-depth knowledge of the model given the APUF variant.

\mypara{Our Contributions} The primary contributions and results of this work are summarized as follows. Significantly, all reported results are achieved with \textit{a common personal computer} and modeling attacks are completed \textit{within an hour} even for large-scaled strong APUF variants.

\begin{itemize}[noitemsep, topsep=2pt, partopsep=0pt,leftmargin=0.4cm]
  \item We are the first to introduce multi-label/head classification to facilitate multi-SCI DL modeling attack, coined as MLMSA that eliminates the curse of dimensionality in the SLMSA. Specifically, the MLMSA model output dimension is now equal to the dimension summation per SCI rather a dimension multiplication per SCI in the SLMSA. In contrast to SLMSA that requires mapping from the predicted label to the response, MLMSA directly outputs the response.
  
  \item We have successfully attacked 128-stage $10$-XOR-APUF, ($2,2,8$)-OAX-APUF and ($5,5$)-iPUF with the MLMSA by simultaneously using the \textit{response and easy-to-obtain reliability SCI}. Notably, 128-stage $12$-XOR-APUF, ($2,2,9$)-OAX-APUF are also breakable statistically, that is, among five repetitions, one attempt succeeds in our experiments. For these attacks, the training size is no more than $600,000$ and training completes within an hour. In contrast to GRA, the MLMSA does not require a mathematical model of underlying PUFs. As a comparison, the purely CRP based DL modeling attacks can break 128-stage $7$-XOR-APUF but with significantly increased training size of 30M \cite{wisiol2021neural}.
  
\item We have advanced the breakable APUF variants to a even larger scale, albeit the concise design of the proposed MLMSA. By simultaneously exploiting multiple SCIs including response, power and reliability, the MLMSA successfully breaks $30$-XOR-APUF, ($2,2,30$)-OAX-APUF,
($9,9$)- and ($2,18$)-iPUFs, all with 128-stage underlying APUFs.
  
\item Based on silicon measurements, we have further affirmed the merits of leveraging \textit{additional easy-to-obtain reliability information} to attack XOR-APUFs compared to the setting of merely using response information. In particular, the response and reliability based DL can successfully attack 128-stage 10-XOR-APUF with 1.5M challenges corresponded response and reliability pairs,
whereas merely response based DL can only attack a 6-XOR-APUF with the same 128-stage and training set size. 

\end{itemize}

\section{Background}\label{sec:background}
This section provides necessary background on APUF and its representative variants that this study examines.

\subsection{Arbiter-based PUF}
The APUF exploits manufacturing variability that results in random interconnect and transistor gate time delays~\cite{gassend2002silicon}. This structure is simple, compact, and capable of yielding a large CRP space. 
In contrast to the optical PUF that lacks mathematical model~\cite{pappu2002physical}, the APUF has a linear additive structure, leading to vulnerability to modeling attacks. In the modeling attack, an attacker utilizes observed CRPs to build a mathematical PUF model that can accurately predict responses for unseen challenges~\cite{ruhrmair2010modeling,ruhrmair2013puf,becker2015gap,becker2015pitfalls}. 

\mypara{$\textbf{Linear Additive Delay Model}$}A linear additive delay model of APUFs is formulated as~\cite{lim2005extracting}:
  \begin{equation}
   \Delta = \boldsymbol{w}^{\rm T}\boldsymbol{\Phi} \label{con:delta},
  \end{equation}
where $\boldsymbol{w}$ is the weight vector that characterizes the time delay segments in the APUF, and $\boldsymbol{\Phi}$ is the parity (or
feature) vector that can be generally understood as a transformation of the challenge. The dimension of both  $\boldsymbol{w}$ and $\boldsymbol{\Phi}$ is $n+1$ given an $n$-stage APUF, where.
 \begin{equation}
     \boldsymbol{\Phi}[n]=1,\boldsymbol{\Phi}[i]=\prod_{j=i}^{n-1}{(1-2\boldsymbol{c}[j])},i=0,...,n-1. \label{con:feature vector}
  \end{equation}

The response of an $n$-stage APUF is determined by the delay difference $\Delta$ between the top path and bottom path of the APUF. This delay difference is the sum of the delay differences of each individual $n$ stages. The delay difference of each stage depends on the corresponding challenge\cite{becker2015gap}. Based on Eq.~\ref{con:delta}, the response $r$ of the challenge $\boldsymbol{c}$ is modeled as:
 \begin{equation}
    r=\begin{cases}
     1, \text{ if } \Delta<0 \\ 
     0, \text{otherwise} .
 \end{cases}
 \label{con:responses}
 \end{equation}

\begin{figure}
	\centering
	\includegraphics[trim=0 0 0 0,clip,width=0.4\textwidth]{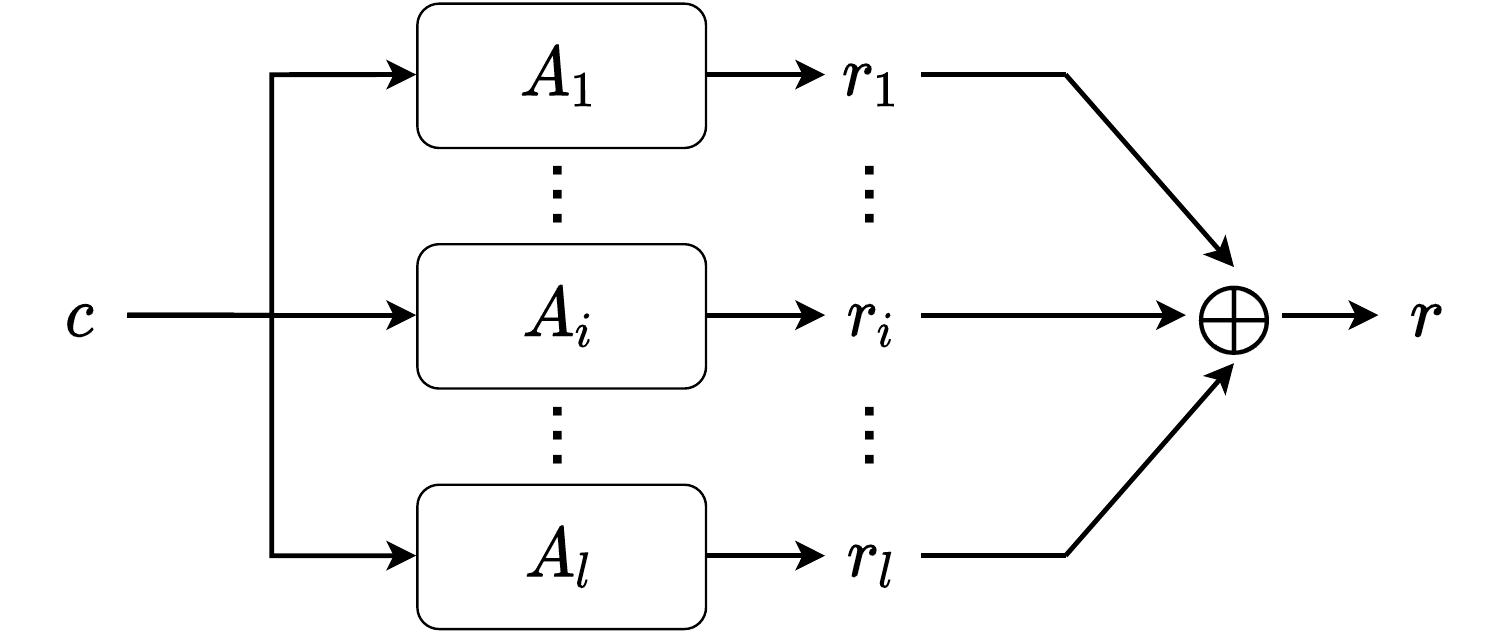}
	\caption{$l$-XOR-APUF consists of $l$ APUFs and each of the APUF response, $\{r_1,...,r_l\}$ is XOR-ed at the end to form a 1-bit response $r$. All APUFs share the same challenge $\boldsymbol{c}$.}
	\label{fig:xor}
\end{figure}
\subsection{XOR-APUF}
As shown in Fig.~\ref{fig:xor}, $l$-XOR-APUF has $l$ underlying APUFs in parallel. Each APUF shares the same challenge and produces a digital response. All $l$ responses are XOR-ed to form the final $l$-XOR-APUF response. Using a larger $l$ can nearly exponentially increase the modeling attack complexity when \textit{only CRP is used}. However, the $l$-XOR-APUF unreliaiblity increases when $l$ is increasing, which negatively restricts the large $l$ usage to some extent. In addition, the large $l$ of a $l$-XOR-APUF is still ineffective against reliability-based modeling attacks---it uses reliability information of the CRP---since the complexity of such attack is only linearly increased as a function of $l$. 

\subsection{OAX-APUF}
\begin{figure}
	\centering
	\includegraphics[trim=0 0 0 0,clip,width=0.4\textwidth]{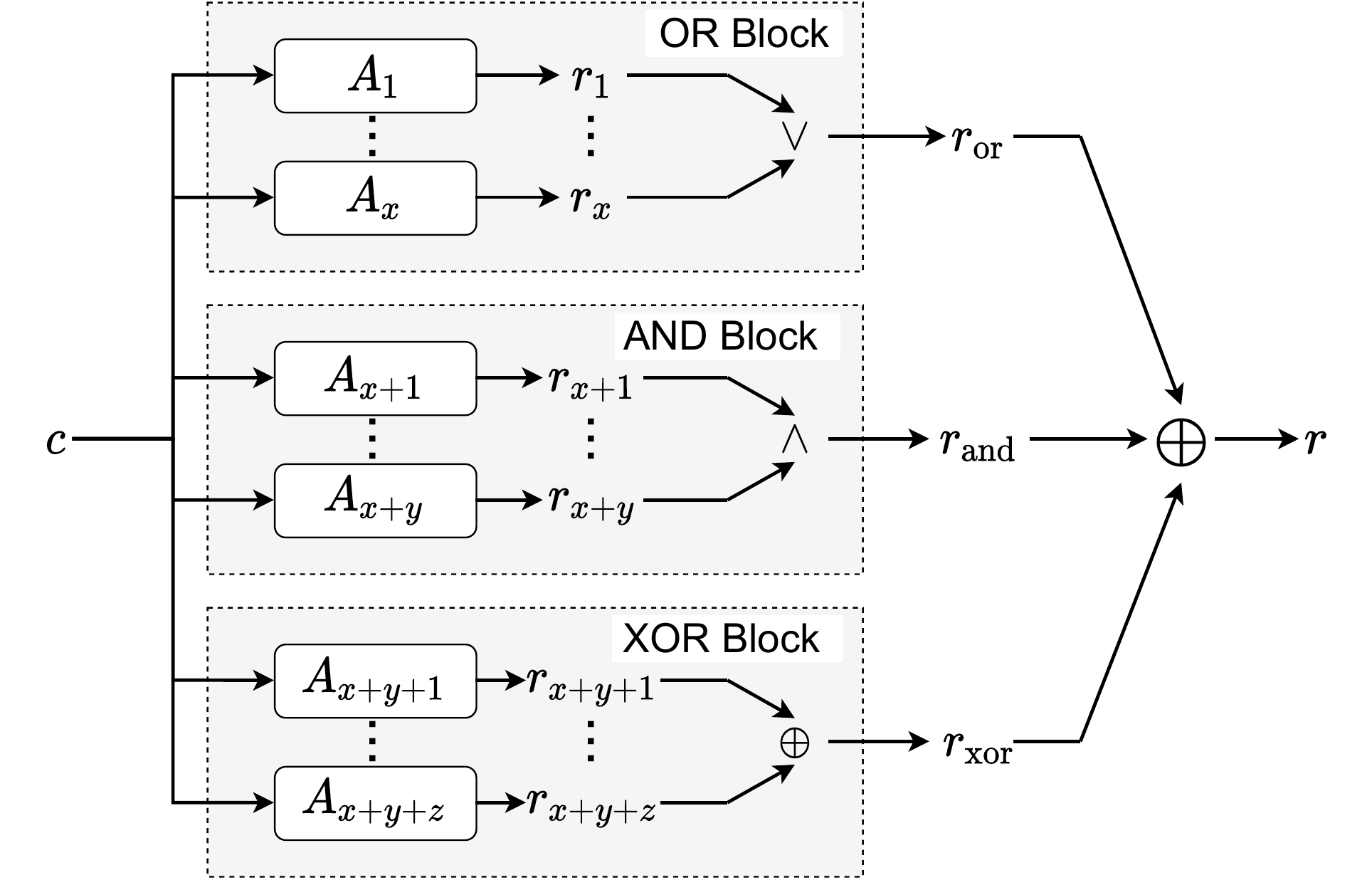}
	\caption{Overview of ($x,y,z$)-OAX-PUF, which has three blocks: OR, AND, and XOR blocks. 
	}
	\label{fig:oax}
\end{figure}

As is shown in Fig.~\ref{fig:oax}, the OAX-APUF~\cite{yao2022design} consists of OR, AND and XOR blocks. The $x$ APUFs' responses are OR-ed to get $r_{\rm or}$, $y$ APUFs are belong to AND block, in which the responses are AND-ed. The XOR block contains $z$ APUFs, whose responses are XOR-ed to gain $r_{\rm xor}$. The responses of three blocks are XOR-ed to obtain the final response $r$. According to~\cite{yao2022design}, the OAX-APUF has higher reliability than XOR-APUF, while OAX-APUF can defeat Covariance Matrix Adaptation Evolution Strategy (CMA-ES) based reliability attacks and demonstrate comparable modeling resilience to logistics regression attack compared to ($x+y+z$)-XOR-APUF. However, it has relatively lower resilience than ($x+y+z$)-XOR-APUF when the DL attack is applied~\cite{yao2022design}.
 
\subsection{iPUF}
\begin{figure}[h]
	\centering
	\includegraphics[trim=0 0 0 0,clip,width=0.40\textwidth]{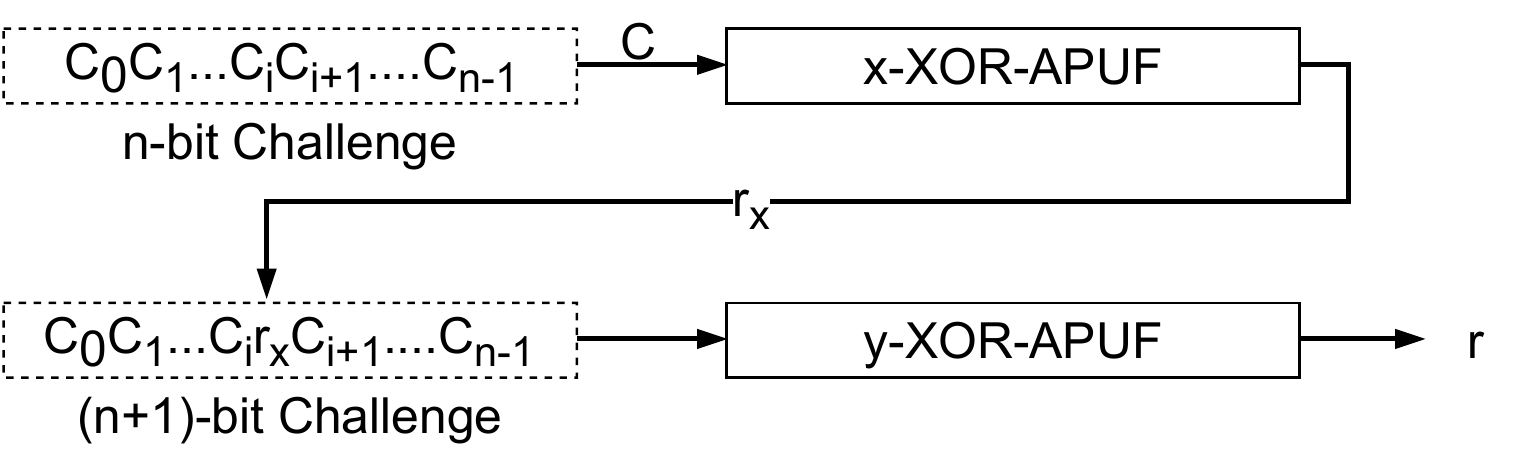}
	\caption{$n$-bit (x,y)-iPUF~\cite{nguyen2019interpose}.}
	\label{fig:ipuf}
\end{figure}

The iPUF contains two layers of XOR-APUFs~\cite{nguyen2019interpose,liu2022multiclass}. As shown in Fig.~\ref{fig:ipuf}, the response of $x$-XOR-APUF is inserted into the $i$-th position of the challenge to obtain the challenge of ($n+1$) bits. The new ($n+1$)-bit challenge is input into $y$-XOR-APUF to get the final response $r$. In theory and experiment, the iPUF has been demonstrated to have desired resistance to LR and reliability based CMA-ES attacks~\cite{nguyen2019interpose,liu2022multiclass}. According to~\cite{wisiol2020splitting}, the security of the ($x,y$)-iPUF against modeling resilience is similar to a ($\frac{x}{2}+y$)-XOR-APUF when the logistic regression (LR)-based divide-and-conquer attack is applied.

\subsection{Feed Forward Arbiter PUF}
\begin{figure}[h]
	\centering
	\includegraphics[trim=0 0 0 0,clip,width=0.45\textwidth]{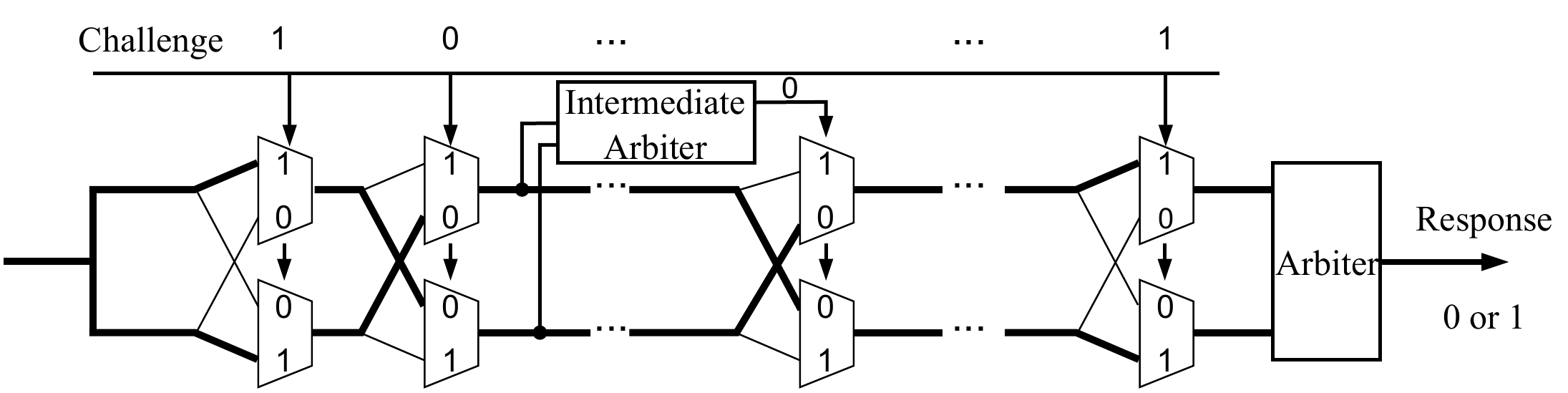}
	\caption{An exemplified FF-APUF with one loop.}
	\label{fig:ff-puf with one loop}
\end{figure}

The Feed Forward Arbiter-PUF (FF-APUF)~\cite{lim2005extracting} adds one or more intermediate arbiters within a basic APUF, and the output response of the intermediate arbiter replaces one or multiple bits of the challenge. This is a typical design of obfuscating the APUF challenge bit(s). The structure of a FF-APUF with one loop is depicted in Fig.~\ref{fig:ff-puf with one loop}~\cite{gao2022systematically}. This FF-APUF can be incorporated with XOR or OAX operations when multiple underlying FF-APUFs are used.

\section{Related Works}\label{sec:relatedwork}
Modeling attacks on strong PUFs normally rely on machine learning (ML) techniques. The ML attacks against strong PUFs can be divided into three categories according to the type of training data used: CRP-based ML attacks, SCI-based attacks and SCI hybrid attacks, which uses CRPs; SCI; and CRPs along with SCI(s) as training data, respectively. 

\subsection{CRP-based ML Attacks}
Logistic regression (LR), support vector machine (SVM), and evolution strategies were utilized by Rührmair~\textit{et el.}~\cite{ruhrmair2010modeling} to model XOR-APUF, FF-APUF and LSPUF in 2010. There are number of improvements to increase the attacking accuracy~\cite{sahoo2015case}. It is always suggested increase the scale of the APUF variants, in particular, the XOR-APUF to increase the modeling resilience against those CRP-based modeling attacks.
In order to increase the complexity of these ML attacks, more APUF variants building upon various forms of recompositions have been proposed, such as MPUF~\cite{sahoo2017multiplexer}, iPUF~\cite{nguyen2019interpose} and OAX-APUF~\cite{yao2022design}.

According to~\cite{nguyen2019interpose}, LR is the most efficient attack against $l$-XOR-APUF, but it cannot be used to attack the iPUF directly. Wisiol~\textit{et el.}~\cite{wisiol2021neural} made some improvements to the LR attack and reported that the improved LR attack can break $64$-bit $8$-XOR-APUF with an accuracy of 96.4\% by using up to $150$M CRPs (i.e., training time is 391 minutes using 4 threads). The LR-based divide-and-conquer attack~\cite{wisiol2020splitting} (LDA) was proposed to attack the iPUF, which can successfully break $64$-stage ($1,7$)-iPUF with an accuracy of 97\%~\cite{liu2022multiclass}. As reported by Liu~\textit{et el.}~\cite{liu2022multiclass}, the LDA attack can successfully break $128$-stage ($6,6$)-iPUF. According to Wisiol~\textit{et el.}~\cite{wisiol2021neural}, the MLP-based divide-and-conquer attack is able to attack 64-stage ($11,11$)-iPUF with $650$M CRPs.

More recently, DL has been shown to be a simple and effective way to attack strong PUFs \textit{without knowing the underling mathematical strong PUF model}. Alkatheriri~\textit{et el.}~\cite{alkatheiri2017towards} showed that a $1$-hidden layer MLP attack can successfully model FF-APUF with $6$ loops in 2017. In 2018, Aseeri~\textit{et el.}~\cite{aseeri2018machine} proposed a $3$-hidden layer MLP attack, which can successfully model $128$-bit $7$-XOR-APUF with $40$M CRPs according to Wisiol~\textit{et el.}~\cite{wisiol2021neural}. Santikellur~\textit{et el.}~\cite{santikellur2019deep} proposed DL attacks on XOR-APUF, MPUF and iPUF in 2019, which can break $128$-stage ($4,4$)-iPUF, ($128,5$)-rMPUF and $5$-XOR-APUF. It has been also shown that the ($64,6$)-rMPUF and ($32,7$)-rMPUF are breakable~\cite{alamro2021machine}. Note compared to iPUF, XOR-APUF and OAX-APUF, it requires greatly increased APUFs and many MUXs (e.g., $2^k$-to-1 MUXs are first decomposed into many $2$-to-1 MUXs for implementation). The implementation of a further scaled ($64,7$)-rMPUF and ($32,8$)-rMPUF requires at least 255 64-stage APUFs and 511 32-stage APUFs, respectively, which high area overhead renders its practicality to a large extent.
Mursi~\textit{et el.}~\cite{mursi2020fast} proposed a 3-hidden layer MLP attack, which mainly focuses on XOR-APUF. 
According to Wisiol~\textit{et el.}~\cite{wisiol2021neural}, this 3-hidden layer MLP~\cite{mursi2020fast} can successfully break 128-stage $7$-XOR-APUF with $30$M fully reliable CRPs.

\subsection{SCI-based Attacks}
SCI-based attacks can be divided into pure side-channel analysis (SCA) attacks and SCA-based ML attacks. The pure SCA attacks can be conducted alone to attack a single APUF, such as reliability-based analysis~\cite{delvaux2013side} and the photonic emission attack~\cite{tajik2017photonic}. The SCA-based ML attacks mainly utilize reliability and power SCIs.

The reliability-based ML attack establishes the reliability model of PUF that exploits the relationship between the response reliability and internal parameters~\cite{becker2015gap}. The measured reliability data and challenges are provided to, e.g., CMA-ES model, as training data to learn the internal parameters of e.g., XOR-APUFs. The fault injection can be utilized to accelerate the reliability SCI collection~\cite{delvaux2014fault,liu2022multiclass}. The reliability-based CMA-ES attack~\cite{becker2015gap} can successfully break XOR-APUF and LSPUF. The CMA-ES attack is based on the assumption that the unreliability contribution per APUF of the $l$-XOR-APUF is equal, so that the CMA-ES can converge to any of $l$ APUFs in an equal chance when the attack repeats. Therefore, the complexity of breaking the $l$-XOR-APUF is linear in $l$. The OAX-APUF~\cite{yao2022design} and iPUF~\cite{nguyen2019interpose} breaks such an assumption, thus can defeat the CMA-ES based reliability modeling attacks.

Different power-based ML attacks leverage differing methods for analyzing power leakages, e.g., simple power analysis (SPA) and correlation power analysis (CPA)~\cite{liu2022multiclass}. Becker~\textit{et el.}~\cite{becker2014active} proposed a CPA-based CMA-ES attack that uses power correlation coefficients as the fitness function to model controlled PUFs and LSPUFs. An SPA-based LR attack was proposed by Rührmair~\textit{et el.}~\cite{ruhrmair2014efficient}, which adopts a gradient-based algorithm similar to LR to learn the power side-channel model of XOR-APUF. However, because the relationship between other APUF variants' power and response is difficult to deduce, so fewer power-based ML attacks are used to model other APUF variants.

Though there are a number of SCI sources that can be used to attack strong PUFs, the reliability SCI is the most easily obtainable one. To collect power SCI, physical access to the PUF device and some expertise are required. The photonic emission collection is costly and usually requires proficient expertise.

\subsection{SCI Hybrid Attacks}
The above two types of ML based attacks with a CRP-based attack or SCI-based attack only use CRP or SCI, and thus knowledge gained by both is not utilized. The other type of SCI hybrid ML attack considers using them concurrently to be more efficient. There are two recent studies on SCI hybrid attacks, exhibiting greatly improved attack efficacy. One is the gradient-based reliability hybrid attack (GRA)~\cite{tobisch2021combining} and the other is the multi-class/single-label multi-SCI attack (SLMSA)~\cite{liu2022multiclass}. 

The GRA~\cite{tobisch2021combining} was mainly devised to attack ($x,y$)-iPUF. In essence, it combines the CMA-ES reliability attack and LR CRP attack together. For the CMA-ES reliability attack term, it learns multiple APUFs concurrently (i.e., regular CMA-ES learn an APUF per run~\cite{becker2015gap}) and enforces that each APUF is dissimilar to others to prevent the APUF converging to those easiest-to-learn APUFs through reliability SCI. 
The GRA attack requires careful constraints to each attack term, which potentially requires manual settings in practical upon trials. Note that the GRA attack is less effective on ($x,y$)-iPUF with $x>1$. Therefore, a multiple pass attack similar to the iPUF splitting attack~\cite{wisiol2020splitting} has to be adopted. In this context, the $y$-APUF are firstly learned, then the $x$-APUF are learned sequentially. In addition, the GRA requires to construct a differential model for the iPUF, which is non-trivial for adoption as it requires in-depth understanding of the underlying PUFs under attack.

The multi-class classification based side-channel hybrid attack (SLMSA) proposed by Liu \textit{et al.}~\cite{liu2022multiclass} is the state-of-the-art to attack XOR-APUF and iPUF, which avoids the underlying PUF mathematical models by using DL techniques. The SLMSA combines response and power SCI to validate its efficiency. To transform the hybrid information into multiple classes, where there is only one true value and the rest are false values, we use one-hot vector encoding also referred to as single-label classification, so the SLMSA has to firstly fuse CRP information with SCI to construct the so called challenge-synthetic-feature pairs (CSPs) via the feature crossing method of Liu {\it et al.}~\cite{liu2022multiclass}. 
More specifically, feature crossing uses the Cartesian product of the response \textbf{r} and side-channel information \textbf{p} to form a single label. Therefore, the number of categories of the multi-class classification is substantially increased when the dimension per \textbf{p} and number of \textbf{p} increases due to usage of the Cartesian product. 
This could incur dimension curse as recognized by Liu \textit{et al.}~\cite{liu2022multiclass}. For example, the response has two categories, and the $l$-XOR-APUF has ($l+1$) power SCI categories, so that the dimension is $2\times (l+1)$.
In this context, Liu \textit{et al.} take the CRP and the power SCI into consideration. The reliability SCI is not used. Nonetheless, the SLMSA has shown to successfully break $128$-stage $16$-XOR-APUF and $(2,16)$-iPUF with $600,000$ training CSPs (response and power SCI). However, this study does not validate the efficacy of the reliability SCI that is the most easily obtainable SCI. 
In addition, the prediction of the response is not immediately available, which is recovered through a remapping according to the CSP process.

\section{Multi-Label Multi-SCI based Deep Learning Attacks}\label{sec: multilabel classification}

We propose a multi-label DL based attack to efficiently and effectively take advantage of multiple SCIs, coined as MLMSA. Firstly, its output dimension is merely a summation per used SCI. Secondly, its can directly predict the response of the learned strong PUF without additional remapping. Thirdly, it can allow flexible weight tuning per SCIs and response to gain improved attack accuracy---we thus can break 128-stage 12-XOR-APUFs by using response and the \textit{easy-to-obtain reliability SCI} that is not considered in~\cite{liu2022multiclass}.

\subsection{Multi-Label Model for MLMSA}
Liu \textit{et al.} used a single-label model~\cite{liu2022multiclass} to exploit multiple information, e.g., CRP and power SCI. In the single-label model, a given input instance can only belong to one of more than two classes. This results in the inconvenient CSP synthesis, where the dimension of the output is greatly increased, especially when multiple SCIs are concurrently exploited---the {\it curse of dimensionality} recognized by Liu \textit{et al.}~\cite{liu2022multiclass}. We note that the single-label model can be circumvented via the multi-label model. In the multi-label classification, there is no constraint on how many of the classes the instance can be assigned to. For instance, an analogy is that a movie can has multiple classes of comedy, romance, and action in the multi-label output.

The multi-label DL model~\cite{tsoumakas2007multi,yeh2017learning} is also usually referred to as multi-head/output DL model. Different from the single-label/head DL model, the output layer of the multi-head model has multiple outputs or heads, which each head corresponds to a label (i.e., response or power SCI or reliability SCI). Supposing there are $k$ heads/outputs in the multi-head model, then the loss of the model can be expressed as Eq.\ref{eq:total loss}: 

\begin{equation}
\label{eq:total loss}
   L=\sum_{i=1}^{k}\lambda_{i}L_{i}, 
\end{equation}
where $L_{i}$ means the loss of $i$-th head, $\lambda_i$ means the weight or the regularized factor of the $L_{i}$, which can be flexibly tuned to gain optimal performance. 

\begin{figure}[h]
	\centering
	\includegraphics[trim=0 0 0 0,clip,width=0.30\textwidth]{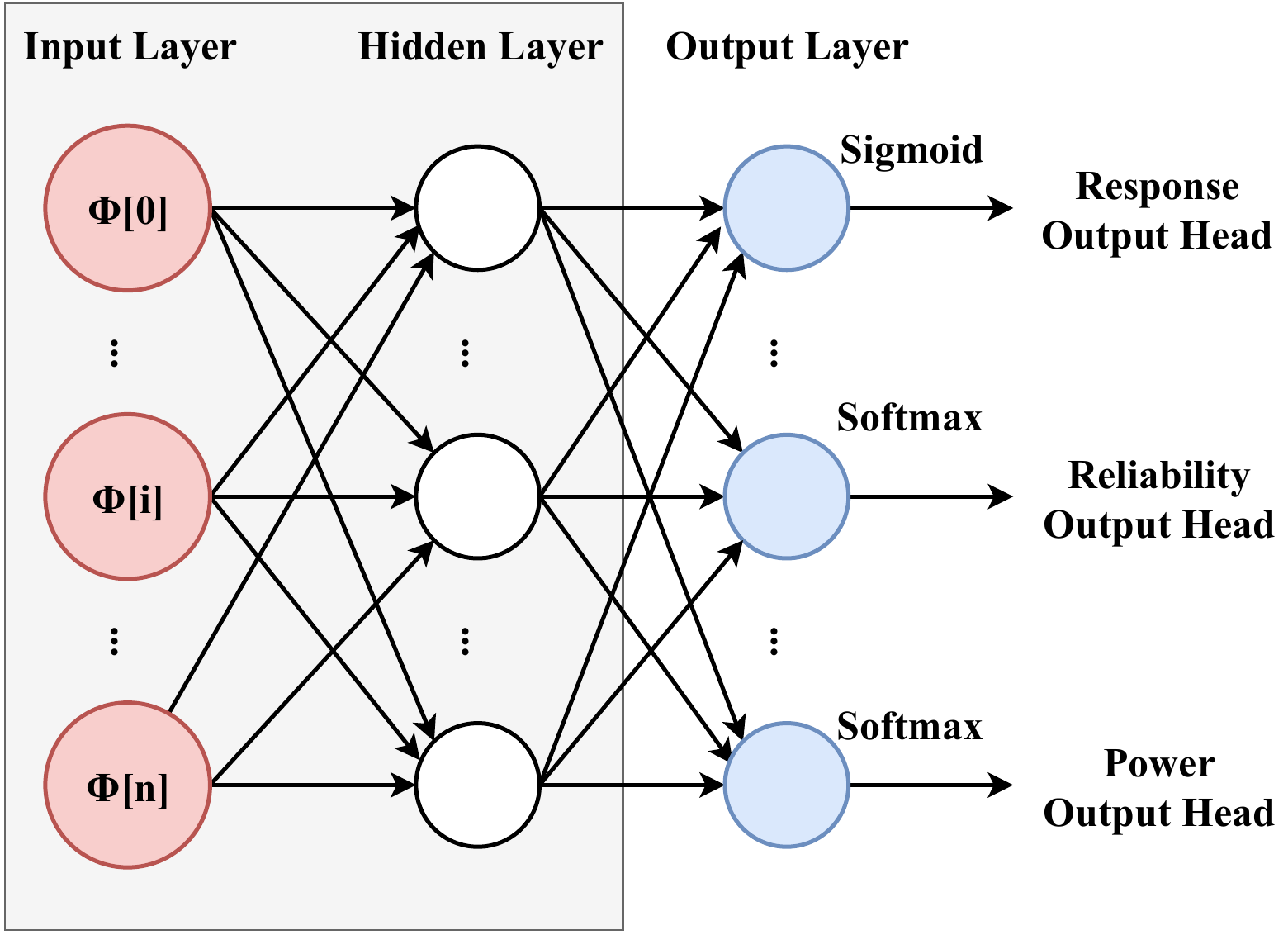}
	\caption{Exemplified three-head DL architecture with heads of response, reliability SCI, and power SCI. The input layer and hidden layer(s) are shared by multiple heads.}
	\label{fig:multi head}
\end{figure}

\subsection{MLMSA}
As shown in Fig.~\ref{fig:multi head}, the multi-head model can use CRP and a number of SCIs, thus enhancing the model for learning the underlying PUFs better by leveraging more useful information sources. Because not only the response but also other SCIs observed for a given challenges are all simultaneously used to train the model, providing more meaningful information to model the internal parameters of the underlying strong PUF. 

This work focuses on power or/and reliability SCIs. The power consumed by the e.g., $l$-XOR-APUF is linearly proportional to number of responses being `1' in $l$ APUFs. More specifically, reliability SCI is obtained by computing the number of responses of `1's from $m$ repeated measurements given the same challenge queries. Let us consider ten repeated measurements of a given challenge as an example: the number of responses with `1's obtained from $10$ repeated measurements has a value ranging from $0$ to $10$, and thus there are $11$ possible values. If the reliability SCI is divided into $11$ categories, each integer number stands for one category. The proposed MLMSA attack has three stages:

\textbf{Pretreatment Stage:} Collecting CRPs and the exploited SCI(s). Note that the label of a given SCI needs to be converted to one-hot vector.

\textbf{Training Stage:} Using multi-head model to train the targeted strong PUF model. The input is a challenge. One head predicts the response, and the other head(s) predict(s) the rest SCI(s), respectively, of the given challenge. The difference between the predictions and the ground-truth labels are used to optimize the multi-head model, according to Eq.~\ref{eq:total loss}.

\textbf{Prediction Stage:} Once the multi-head model is trained, the response given an unseen challenge can be \textit{directly predicted by the response head/output}.

\section{Experimental Results and Analysis}\label{sec:experimental results}

\subsection{Experimental Setup}

According to the dynamic power analysis of PUFs, the amount of drawn charge is linearly proportional to the number of latches exhibiting a value of `1's~\cite{liu2022multiclass}. For PUF designs that employ more than one APUFs in parallel, by measuring the amount of current drawn from the supply voltage during any latch transition, the cumulative number of APUFs that respond with `1's can be determined~\cite{mahmoud2013combined,liu2022multiclass}. 
In other words, the power consumption is linear with the number of `1' responses produced by APUFs. This has been validated by the consistency between physical measurements and simulations~\cite{becker2014active,liu2017combined,ruhrmair2014efficient}. Following~\cite{liu2022multiclass}, we adopt the counted `1's as the power SCI. As for reliability information, we apply the same challenge repeatedly many times, and classify the reliability information according to the number of responses `1's obtained by repeated measurements. More specifically, if there are 10 repeated measurements, the number of categories of reliability SCI is 11 (i.e., from 0 to 10).

Following~\cite{ruhrmair2010modeling,becker2015pitfalls,nguyen2019interpose,wisiol2020splitting,yao2022design}, we use MATLAB to numerically simulate CRPs, power SCI and reliability SCI required by the following experiments---silicon measurement validations of reliability SCI are detailed in Section~\ref{sec:silicon}. Each APUF is $128$ stage---majority of previous studies using 64-stage APUF. For the response and power SCI, most experiments use noise free simulation, in which $\mu= 0$, $\sigma = 1$ are used to generate the weights corresponding to the APUFs as in Eq.~\ref{con:delta}.
In this context, we collect training/testing CRPs. The unreliability is produced by injecting Gaussian noise into the above generated weights by setting, $\mu_{\rm noise} = 0$ and $\sigma_{\rm noise} = 0.05$ to get the noisy weights per repeated same challenge query. The unreliability of APUFs ranges from $0.05$ to $0.08$ after noise injection. The reliability SCI consequentially can be collected. For the power SCI, we count the number of `1's in simulated APUFs as the SCI.

The training set, validation set and test set are divided according to the ratio of 4:1:1. It should be noted that if CRPs used for testing and CRPs used for training collected under different conditions (e.g., enrolled at 25\textcelsius ~but regenerated at 50\textcelsius), testing accuracy is expected to be degraded. 

For FF-APUF, we have only considered the combination of response and reliability, the responses of training set and validation set are obtained by majority voting, and the response of testing set is noise-free. For the XOR-APUF, iPUF, and OAX-APUF, we have considered the hybrid of response, power, and reliability.

The number of hidden layers of the multi-head model as exemplified in Fig.~\ref{fig:multi head} in each experiment is 3 or 4, and the activation function is \textsf{ReLU}. In~\cite{liu2022multiclass}, Liu \textit{et al.} used 2 or 3 hidden layers, which breaks 16-XOR-APUF. Our reproduced results of Liu \textit{et al.} successfully attack 30-XOR-APUF, which can be potentially attributed to the adopted DL architecture with more hidden layers that are optimal values after hyperparameter tuning.
For the response head, the loss function uses \textsf{binary\_crossentropy}, while the loss function of other heads uses \textsf{categorical
\_crossentropy}. The \textsf{Adam} optimizer is used for all experiments. The settings of different head loss weights of MLMSA are summarized in Table~\ref{tab:Loss Weight}. All experiments are completed using a common personal computer with an Intel(R) Core(TM) i5-6200U CPU, and 12~GB memory.

\subsection{Modeling Attacks and Results Analysis}

\begin{table}[]
\caption{MLMSA Multi-Head Loss Weight $\lambda$ Settings.}
\label{tab:Loss Weight}
\centering
\resizebox{0.35\textwidth}{!}{
\begin{threeparttable}
\begin{tabular}{|c|c|c|c|c|}
\hline
\textbf{PUF}              & \textbf{Multi-Head} & \begin{tabular}[c]{@{}c@{}} Response \\ Weight\end{tabular} & \begin{tabular}[c]{@{}c@{}} Power \\ Weight\end{tabular} & \begin{tabular}[c]{@{}c@{}} Reliability \\ Weight\end{tabular} \\ \hline
\multirow{3}{*}{XOR-APUF} & Two-Head A          & 10                       & 2                     & /                           \\ \cline{2-5} 
                          & Two-Head B          & 1                        & /                     & 0.8(1.8)                    \\ \cline{2-5} 
                          & Three-Head          & 10                       & 2                     & 2(1)                        \\ \hline
\multirow{3}{*}{OAX-APUF} & Two-Head A          & 10                       & 2                     & /                           \\ \cline{2-5} 
                          & Two-Head B          & 1                        & /                     & 0.8(1.8)                    \\ \cline{2-5} 
                          & Three-Head          & 10                       & 2                     & 2                           \\ \hline
\multirow{3}{*}{iPUF}     & Two-Head A          & 10(2)                    & 2(3)                  & /                           \\ \cline{2-5} 
                          & Two-Head B          & 1                        & /                     & 0.8                      \\ \cline{2-5} 
                          & Three-Head          & 10(2)                    & 2                     & 2                           \\ \hline
\end{tabular}
\begin{tablenotes}
      \footnotesize
      \item [*]Two-Head A of MLMSA uses response and power SCI. Two-Head B of MLMSA uses response and reliability SCI. Three-Head of MLMSA uses response, power and reliability SCIs.
      \item [**] In Two-Head B, for $l$-XOR-APUF, the reliability weight is $1.8$ when $l=10$; and $0.8$ in other cases. In Three-Head, the reliability weight is $1$ when $l=29, 30$; $2$ in other cases.
      \item [***] In Two-Head B, for ($x,y,z$)-OAX-APUF, the reliability head loss weight is $1.8$ when $x+y+z=12$; $0.8$ in other cases.
      \item [****] In Two-Head A, for ($8,8$)-iPUF, ($9,9$)-iPUF, ($2,16$)-iPUF and ($2,18$)-iPUF, the response weight is $2$, power weight is $3$, these two weights are $10$ and $2$ respectively in other cases. In Three-Head, for ($8,8$)-iPUF, ($9,9$)-iPUF, ($2,16$)-iPUF and ($2,18$)-iPUF, the response weight is $2$. In Three-Head, for other iPUFs, the weight of response is $10$ (Three-Head).
      \item [*****] These head loss weight settings are based on few empirically tuning. There may be better choices, which need to be analyzed on a case by case basis.

\end{tablenotes}
\end{threeparttable}
}
\end{table}

The XOR-APUF, OAX-APUF, iPUF, and FF-APUF are used to validate the effectiveness of the proposed MLMSA. 
Note that the SLMSA is the most efficient attack using not only CRPs but also SCI (in particular, power SCI).
We compare the results with SLMSA by reproducing it under the same experimental settings for fair comparisons.
Table~\ref{tab: attack results} summarizes the main results of the MLMSA and SLMSA attacks on three strong APUF variants. Notably, the Multi-Class A and Multi-Class B belong to the SLMSA attacks~\cite{liu2022multiclass}, where different SCIs are used:
\begin{enumerate}
    \item \textbf{Multi-Class A:} The CSPs are formed with power SCI and CRPs;
    \item \textbf{Multi-Class B:} The CSPs are formed with reliability SCI and CRPs.
\end{enumerate}

Note that the Multi-Class B is not considered in~\cite{liu2022multiclass} for experimental evaluations, which we explore, for the first time, for comparison purposes.

\begin{table*}[]
\caption{Comparisons of MLMSA with SLMSA (i.e., multi-class)~\cite{liu2022multiclass} and DL with pure CRP attack~\cite{santikellur2019deep} on complicated strong PUFs.}
\label{tab: attack results}
\centering
\resizebox{0.8\textwidth}{!}{
\begin{threeparttable}
\begin{tabular}{|c|c|c|c|c|c|c|c|}
\hline
\textbf{PUF}      & \textbf{Training CRPs}        & \textbf{Two-Head A} & \textbf{Three-Head} & \textbf{Multi-Class A~\cite{liu2022multiclass}} & \textbf{Two-Head B} & \textbf{Multi-Class B~\cite{liu2022multiclass}} & \textbf{DL2019~\cite{santikellur2019deep}} \\ \hline
5-XOR-APUF        & 300,000 (600,000 / 655,000)   & 96.97\%             & 97.45\%             & 98.25\%                & 98.51\%             & 98.27\%                & 97.61\%          \\ \hline
6-XOR-APUF        & 300,000 (600,000 / 1,200,000) & 96.98\%             & 97.38\%             & 98.09\%                & 98.24\%             & 97.54\%                & 49.96\%          \\ \hline
10-XOR-APUF       & 300,000 (600,000)             & 95.43\%             & 95.61\%             & 97.12\%                & 96.14\%             & 96.32\%                & /                \\ \hline
30-XOR-APUF       & 600,000                       & 91.85\%             & 89.77\%             & 92.13\%                & /                   & /                      & /                \\ \hline
(2,2,3)-OAX-APUF  & 300,000 (600,000)             & 97.49\%             & 97.94\%             & 97.40\%                & 98.08\%             & 97.87\%                & 97.57\%          \\ \hline
(2,2,4)-OAX-APUF  & 300,000 (600,000)             & 97.18\%             & 97.82\%             & 97.37\%                & 97.98\%             & 96.81\%                & 50.08\%          \\ \hline
(2,2,8)-OAX-APUF  & 300,000 (600,000)             & 96.17\%             & 95.00\%             & 96.31\%                & 95.30\%             & 96.84\%                & /                \\ \hline
(2,2,30)-OAX-APUF & 600,000             & 87.27\%             & 84.23\%             & 88.95\%                & /                   & /                      & /                \\ \hline
(4,4)-iPUF        & 300,000 (600,000 / 647,000)   & 97.05\%             & 97.58\%             & 97.33\%                & 96.82\%             & 96.63\%                & 74.73\%          \\ \hline
(5,5)-iPUF        & 600,000 (1,200,000)           & 96.94\%             & 97.29\%             & 97.09\%                & 95.70\%             & 95.36\%                & /                \\ \hline
(8,8)-iPUF       & 600,000                       & 95.79\%             & 94.35\%             & 95.71\%                & /                   & /                      & /                \\ \hline
(9,9)-iPUF       & 600,000                       & 95.29\%             & 94.72\%             & 95.47\%                & /                   & /                      & /                \\ \hline
(2,16)-iPUF       & 600,000                       & 92.81\%             & 90.33\%             & 92.82\%                & /                   & /                      & /                \\ \hline
(2,18)-iPUF       & 600,000                       & 89.26\%             & 89.00\%             & 90.49\%                & /                   & /                      & /                \\ \hline
\end{tabular}
\begin{tablenotes}
\footnotesize
      \item [*]Two-Head A of MLMSA uses response and power SCI. Two-Head B of MLMSA uses response and reliability SCI. Three-Head of MLMSA uses response, power SCI, and reliability SCI. Multi-Class A of SLMSA uses response and power SCI. Multi-Class B of SLMSA uses response and reliability SCI. 
      \item [**] For Two-Head A, Three-Head and Multi-Class A, the training size is $300,000$ or $600,000$. Take $5$-XOR-APUF as an example, the size of $300,000$ is used when the attacks use power SCI; $600,000$ when attacks use reliability SCI. The $655,000$ is the number given by~\cite{santikellur2019deep}.

\end{tablenotes}
\end{threeparttable}
}
\end{table*}

\subsubsection{$l$-XOR-APUF}
The loss weight settings of multi-head model are described in Table~\ref{tab:Loss Weight}. As for the Two-Head A model (response head and rower head), the response loss weight is $10$, the power loss weight is $2$.
As for Two-Head B model (response head and reliability head), the response weight is $1$, the reliability loss weight is $0.8$ when $l\leq 9$; $1.8$ when $l=10$. As for Three-Head, the response loss weight is $10$, the power loss weight is $2$. While the reliability loss weight is $2$ when $l\leq 28$; $1$ when $l=29,30$.
The training size is $300,000$ when $l\leq 12$; $600,000$ when $l\geq16$ if the attack uses \textit{power} SCI (e.g., Two-Head A, Three-Head and Multi-Class A that is the SLMSA).
As for Two-Head B that only uses the easy-to-obtain reliability SCI, the training size is $600,000$ for all $l$ settings---the largest $l$ in this case is $10$.

\begin{figure}[h]
	\centering
	\includegraphics[trim=0 0 0 0,clip,width=0.35\textwidth]{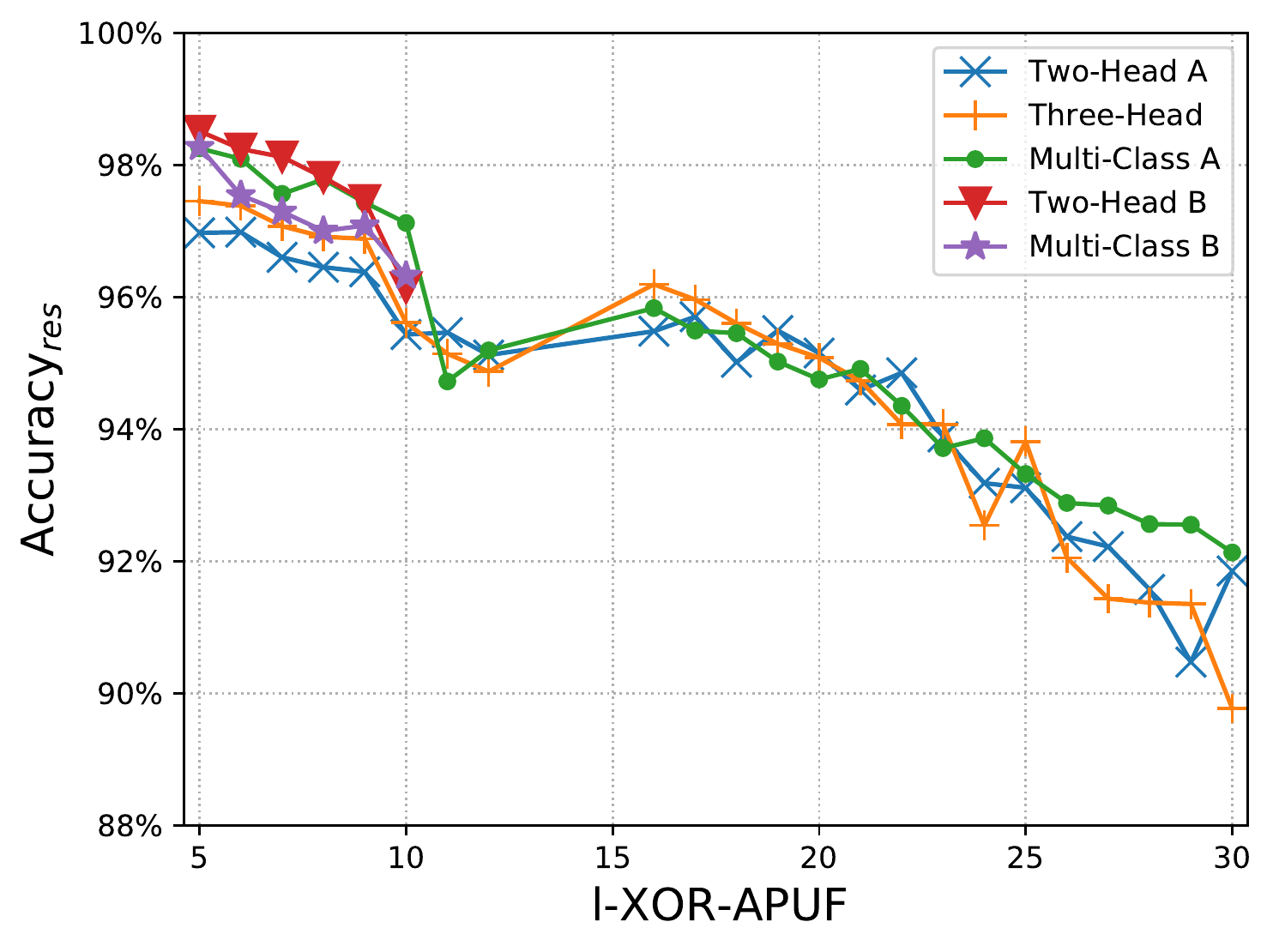}
	\caption{Comparisons of MLMSA (i.e., multi-head) and SLMSA (i.e., Multi-Class) attacks using CRP, or/and power or/and reliability SCI on $l$-XOR-APUFs. The x-axis stands for the $l$ of XOR-APUFs.}
	\label{fig: xor results}
\end{figure}

Fig.~\ref{fig: xor results} depicts the results of multi-head (i.e., our MLMSA) and multi-class (in particular, the SLMSA) attacks on $l$-XOR-APUF ($l\leq 30$). When the power SCI is used, both MLMSA and SLMSA classification can attack 128-stage $30$-XOR-APUF with accuracy about 90\%, which scale has not been achieved in all previous studies. Liu~\textit{et el.}~\cite{liu2022multiclass} only reported the accuracy of 97.8\% when modeling $16$-XOR-APUF by using Multi-Class A---as mentioned above, one potential reason is that Liu~\textit{et el.}~\cite{liu2022multiclass} used 2 or 3 hidden layers that is less powerful than 3 or 4 hidden layers we adopted in the reproduction. 

When using \textit{only response and reliability SCI}, the two attacks can reliably break $10$-XOR-APUF with accuracy more than 95\%---later we show $12$-XOR-APUF is statistically breakable under multiple repeated attacks. The larger $l$, the harder to minimize reliability loss during the training optimization. The Three-Head attack using response, power, and reliability SCI exhibits an improvement over the two-head model that uses response and power SCI only when $l$ is small.
This means the power SCI is more efficient than reliability SCI for attacks.
The accuracy of Two-Head A, Three-Head and Multi-Class~A are similar,
This further indicates the dominant contribution of the power SCI compared to reliability SCI. Notably, \textit{when the power SCI is unavailable}, reliability SCI can indeed help to break larger scale strong PUFs that CRP based modeling attacks cannot achieve alone, as specifically validated in Section~\ref{sec:relattack}.

\subsubsection{($x,y$)-iPUF}
The loss weight settings for the ($x,y$)-iPUF are detailed in Table~\ref{tab:Loss Weight}. For most of the iPUF configurations with 128-stage $x$-APUF and 129-stage $y$-APUF, the response head loss weight is set to $10$, the power head loss weight is $2$, and the reliability head loss weight is $2$. 
However, these settings are not successful in attacking ($8,8$)-iPUF and ($2,16$)-iPUF. It is necessary to tune the corresponding loss weight settings for optimization, the response head loss weight is $2$, the power head loss weight is $3$ when the Two-Head A attack is used; the response head loss weight is $2$ when the Three-Head attack is used. 
As for Two-Head A, Three-Head and Multi-Class A, the training size is $600,000$. While for Two-Head B and Multi-Class B, the training size is $600,000$ for ($4,4$)-iPUF, $1,200,000$ for ($5,5$)-iPUF, respectively.

Though MLMSA is simple, it can also break ($2,16$)/($8,8$)-iPUF with accuracy of 90.33\%/94.35\% that is comparable to the SLMSA when response, power SCI and reliability SCI are exploited. When using response, power SCI and reliability SCI, Three-Head of MLMSA can also successfully model ($2,18$)/($9,9$)-iPUF with accuracy of 89.00\%/94.72\%, which has not been reached by existing works include~\cite{liu2022multiclass}. As for Two-Head B by using response and reliability SCI, both the MLMSA and SLMSA can break ($5,5$)-iPUF with accuracy more than 95\%.
 
\begin{figure}[h]
	\centering
	\includegraphics[trim=0 0 0 0,clip,width=0.35\textwidth]{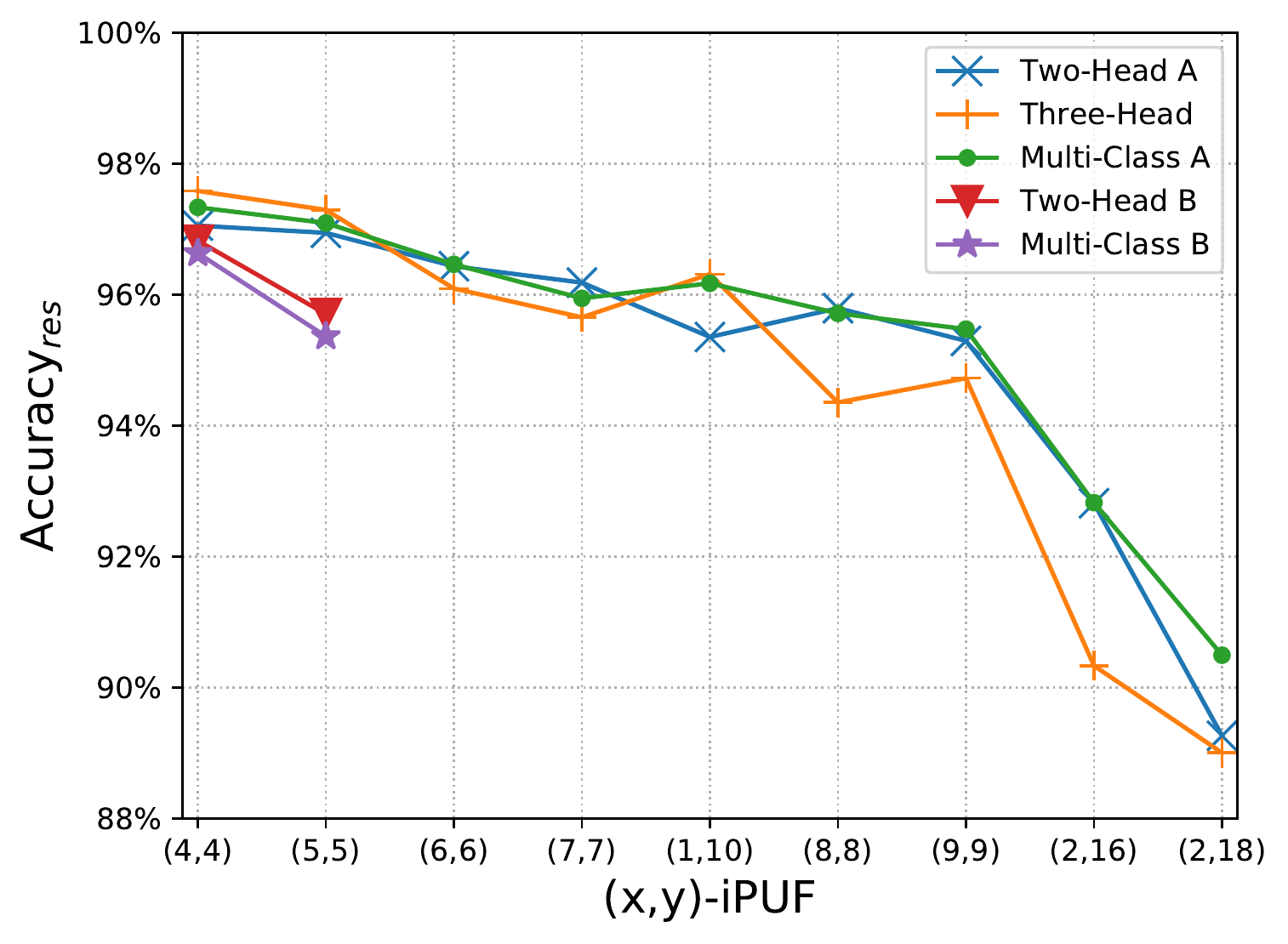}
	\caption{Comparisons of MLMSA (i.e., multi-head) and SLMSA (i.e., Multi-Class) attacks on ($x,y$)-iPUFs. The x-axis stands for the $(x,y)$ of iPUFs.}
	\label{fig: ipuf results}
\end{figure}

\subsubsection{($x,y,z$)-OAX-APUF}
For the ($x,y,z$)-OAX-APUF, we fix $x=2$, $y=2$, and change the setting of $z$. The loss weight settings of multi-head attacks are detailed in Table~\ref{tab:Loss Weight}.
As for Two-Head A model (response head and power head), the response head loss weight is $10$, the power head loss weight is $2$.
As for Two-Head B model (response head and reliability head), the response head loss weight is $1$, the reliability head loss weight is $0.8$ when $z\leq 7$; $1.8$ when $z=8$, respectively. As for Three-Head, the response head loss weight is $10$, the power head loss weight is $2$ and the reliability head loss weight is $2$. The training size is $300,000$ when $z\leq 10$; $600,000$ when $z\geq12$, respectively, if the attacks use power SCI (Two-Head A, Three-Head, and Multi-Class A). As for Two-Head B, the training size is $600,000$.

\begin{figure}[h]
	\centering
	\includegraphics[trim=0 0 0 0,clip,width=0.35\textwidth]{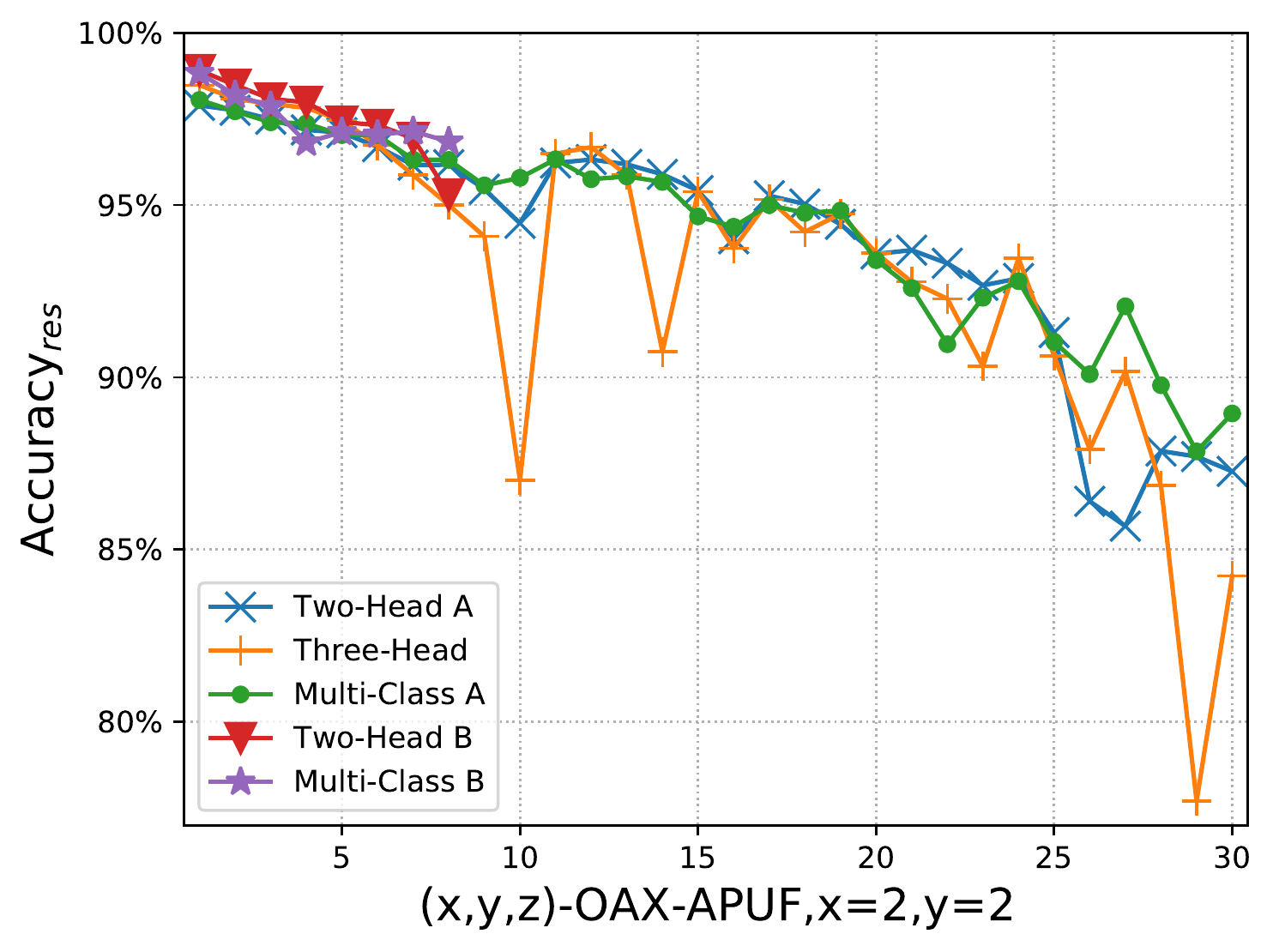}
	\caption{Comparisons of MLMSA and SLMSA using power and reliability SCIs on OAX-APUFs. The x-axis stands for the $z$ of $(x,y,z)$-OAX-APUFs. }
	\label{fig: oax results}
\end{figure}

As shown in Fig.~\ref{fig: oax results}, the performance of MLMSA and SLMSA are similar though the MLMSA is simpler. More specifically, when power SCI is leveraged, both MLMSA and SLMSA can break ($2,2,30$)-OAX-APUF with an accuracy of about 88\%. 
Two-Head B with reliability SCI can reliably break ($2,2,8$)-OAX-APUF with an accuracy of more than 95\%---later in Section~\ref{sec:relattack} we show that ($2,2,9$)-OAX-APUF is statistically breakable. 

These experiments further validate the security of the OAX-APUF. Compared with the $l$-XOR-APUF, the ($x,y,z$)-OAX-APUF with $l=x+y+z$ is slightly easier to be modeled in front of DL based attacks, because the OR and AND are easier to be approximated than the XOR operation by DL. Despite the OAX-APUF defeats CMA-based reliability modeling attacks and improves the modeling resilience to the LR based modeling attacks with only CRPs for training~\cite{yao2022design}.

\subsubsection{FF-APUF}

For FF-APUF, we compare i) the Two-Head model of MLMSA with multi-class of SLMSA attack and ii) the pure CRP based DL attack~\cite{alkatheiri2017towards}. Note for the first two type attacks, only the reliability SCI is utilized.

In this experiment, the number of hidden layer is set to be 2 for Two-Head B and Multi-Class B. The training size is $30,000$ when the loop number is less than $4$; $600,000$ when the loop number is $4$, $5$ and $6$. The weight of response head loss is $10$, and the weight of reliability head loss is $2$. There are three reliability SCI settings: $10$ times of repeated measurement with $11$ classes; $19$ times measurements with $4$ classes (e.g., 0-4 are one class, 5-9 are one class); and $19$ measurements with $20$ classes. The challenge feature vector extraction method is consistent with~\cite{alkatheiri2017towards}.

As results detailed in Table~\ref{tab: ffpuf results}, the multi-head of MLMSA and multi-class of SLMSA attack can successfully model FF-APUF which has $6$ loops with an accuracy of about 90\%. Both attacks that are hybrid attacks exhibit a better accuracy than the purely CRP-based DL modeling attack~\cite{alkatheiri2017towards}. As for the repeated times of reliability SCI, when the response is measured repeatedly for $19$ times and the results are divided into $20$ categories (i.e., more repeated times and fine-grained class), the response accuracy obtained by the two attacks is the highest. This indicates the higher fine-grained reliability SCI, the better.

\begin{figure}[h]
	\centering
	\includegraphics[trim=0 0 0 0,clip,width=0.35\textwidth]{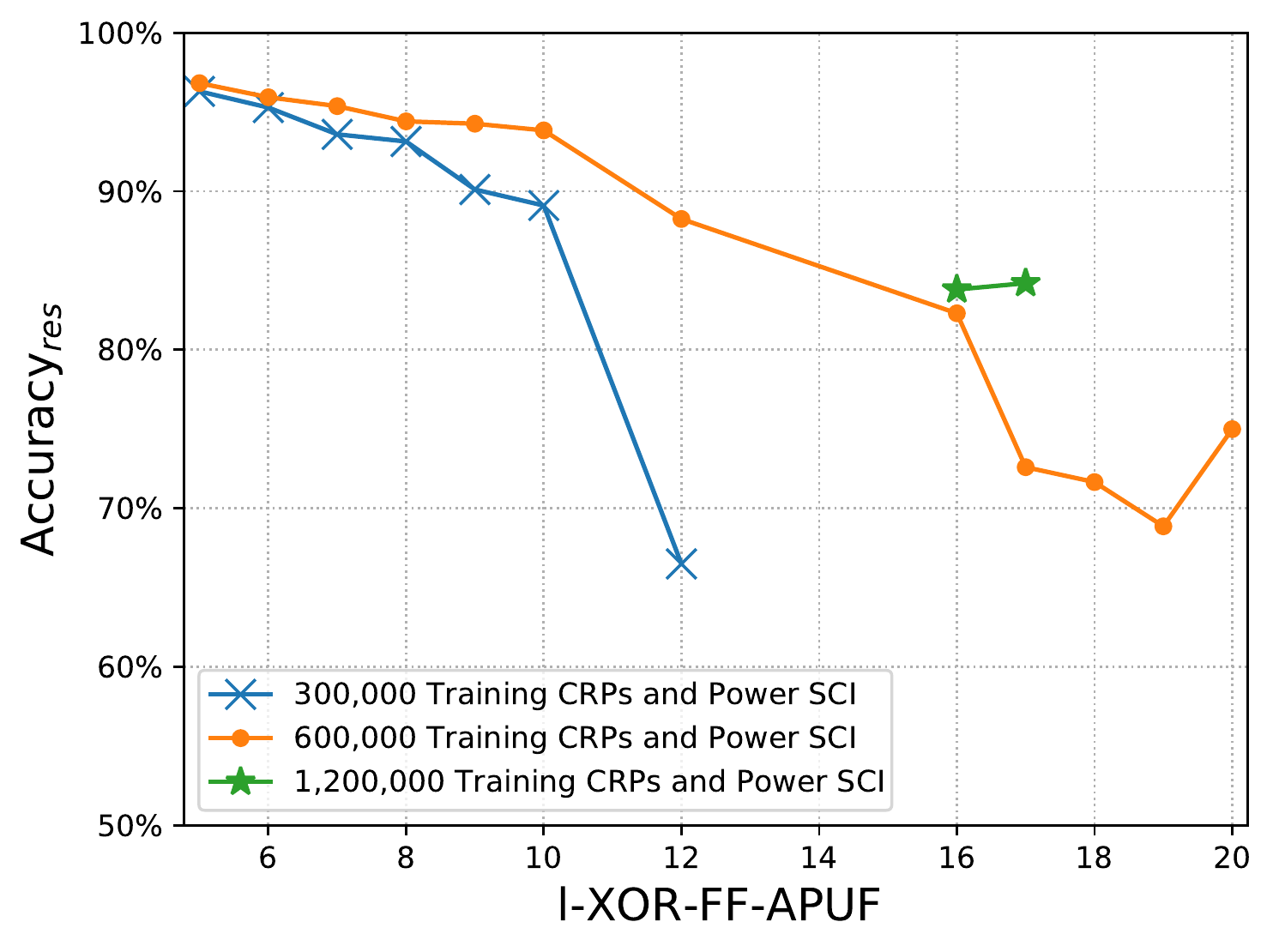}
	\caption{Comparisons of Two-Head A Attack with different number of training CRPs and power SCI on $l$-XOR-FF-APUF in which FF-APUFs have $1$ loop (63→80). The x-axis stands for the $l$ of XOR-FF-APUFs.}
	\label{fig: xorffpuf results}
\end{figure}

MLMSA can be used to break XOR-FF-APUF, where the FF-APUFs are further XOR-ed. As shown in Fig.~\ref{fig: xorffpuf results}, when using response and power SCI, MLMSA can successfully break $10$-XOR-FF-APUF when the training size is $300,000$. By increasing the training size, larger XOR-FF-APUF (i.e., 20-XOR-FF-APUF) is also breakable. Note, here all FF-APUFs have one-loop.

\begin{table*}
\caption{Comparisons of MLMSA with SLMSA (i.e., multi-class)~\cite{liu2022multiclass} and DL with pure CRP attack~\cite{alkatheiri2017towards} on FF-APUFs with varying number of loops.}
\label{tab: ffpuf results}
\centering
\resizebox{0.80\textwidth}{!}{
\begin{threeparttable}
\begin{tabular}{|c|c|c|c|c|c|c|c|c|c|}
\hline
\textbf{loopNum}    & \textbf{($m,cn$)} & \textbf{Start → End}                                                                  & \textbf{Two-Head B} & \textbf{Multi-Class B} & \textbf{TowardsFast} & \textbf{Start → End}                                                                  & \textbf{Two-Head B}            & \textbf{Multi-Class B} & \textbf{TowardsFast} \\ \hline
                    & (19,20)          &                                                                                       & 99.02\%             & 99.04\%                & 95.87\%              &                                                                                       & 98.94\%                        & 99.17\%                & 97.77\%              \\ \cline{2-2} \cline{4-6} \cline{8-10} 
                    & (19,4)           &                                                                                       & 98.75\%             & 99.03\%                & 95.61\%              &                                                                                       & 98.86\%                        & 99.14\%                & 96.85\%              \\ \cline{2-2} \cline{4-6} \cline{8-10} 
\multirow{-3}{*}{1} & (10,11)          & \multirow{-3}{*}{15→80}                                                               & 98.83\%             & 98.91\%                & 95.25\%              & \multirow{-3}{*}{63→80}                                                               & 98.95\%                        & 99.11\%                & 98.06\%              \\ \hline
                    & (19,20)          &                                                                                       & 94.90\%             & 95.00\%                & 95.09\%              &                                                                                       & 94.88\%                        & 95.11\%                & 95.17\%              \\ \cline{2-2} \cline{4-6} \cline{8-10} 
                    & (19,4)           &                                                                                       & 94.79\%             & 94.90\%                & 95.11\%              &                                                                                       & 94.79\%                        & 94.83\%                & 94.99\%              \\ \cline{2-2} \cline{4-6} \cline{8-10} 
\multirow{-3}{*}{2} & (10,11)          & \multirow{-3}{*}{15→80,85}                                                            & 94.82\%             & 94.92\%                & 94.99\%              & \multirow{-3}{*}{63→80,85}                                                            & 94.83\%                        & 94.98\%                & 95.07\%              \\ \hline
                    & (19,20)          &                                                                                       & 94.33\%             & 94.31\%                & 91.84\%              &                                                                                       & 94.34\%                        & 94.59\%                & 93.36\%              \\ \cline{2-2} \cline{4-6} \cline{8-10} 
                    & (19,4)           &                                                                                       & 94.17\%             & 94.24\%                & 91.53\%              &                                                                                       & 94.14\%                        & 94.39\%                & 93.30\%              \\ \cline{2-2} \cline{4-6} \cline{8-10} 
\multirow{-3}{*}{3} & (10,11)          & \multirow{-3}{*}{15→80,85,90}                                                         & 94.26\%             & 94.69\%                & 91.83\%              & \multirow{-3}{*}{63→80,85,90}                                                         & 94.22\%                        & 94.43\%                & 92.88\%              \\ \hline
                    & (19,20)          &                                                                                       & 91.45\%             & 91.53\%                & 91.48\%              &                                                                                       & 91.50\%                        & 91.51\%                & 91.44\%              \\ \cline{2-2} \cline{4-6} \cline{8-10} 
                    & (19,4)           &                                                                                       & 91.41\%             & 91.55\%                & 91.40\%              &                                                                                       & 91.43\%                        & 91.49\%                & 91.59\%              \\ \cline{2-2} \cline{4-6} \cline{8-10} 
\multirow{-3}{*}{4} & (10,11)          & \multirow{-3}{*}{\begin{tabular}[c]{@{}c@{}}15→80,85\\ 15→90,95\end{tabular}}         & 91.43\%             & 91.55\%                & 91.32\%              & \multirow{-3}{*}{\begin{tabular}[c]{@{}c@{}}63→80,85\\ 63→90,95\end{tabular}}         & 91.29\%                        & 91.38\%                & 91.38\%              \\ \hline
                    & (19,20)          &                                                                                       & 90.83\%             & 90.73\%                & 88.62\%              &                                                                                       & 91.12\%                        & 91.04\%                & 89.62\%              \\ \cline{2-2} \cline{4-6} \cline{8-10} 
                    & (19,4)           &                                                                                       & 90.66\%             & 90.80\%                & 88.69\%              &                                                                                       & 90.93\%                        & 91.05\%                & 89.39\%              \\ \cline{2-2} \cline{4-6} \cline{8-10} 
\multirow{-3}{*}{5} & (10,11)          & \multirow{-3}{*}{\begin{tabular}[c]{@{}c@{}}15→80,85\\ 15→90,95,100\end{tabular}}     & 91.00\%             & 91.01\%                & 88.21\%              & \multirow{-3}{*}{\begin{tabular}[c]{@{}c@{}}63→80,85\\ 63→90,95,100\end{tabular}}     & 91.12\%                        & 90.91\%                & 89.49\%              \\ \hline
                    & (19,20)          &                                                                                       & 91.28\%             & 91.38\%                & 89.69\%              &                                                                                       & 91.22\%                        & 91.30\%                & 90.15\%              \\ \cline{2-2} \cline{4-6} \cline{8-10} 
                    & (19,4)           &                                                                                       & 91.33\%             & 91.17\%                & 89.52\%              &                                                                                       & {\color[HTML]{080808} 90.92\%} & 91.21\%                & 89.67\%              \\ \cline{2-2} \cline{4-6} \cline{8-10} 
\multirow{-3}{*}{6} & (10,11)          & \multirow{-3}{*}{\begin{tabular}[c]{@{}c@{}}15→80,85,90\\ 15→95,100,105\end{tabular}} & 91.02\%             & 91.21\%                & 89.69\%              & \multirow{-3}{*}{\begin{tabular}[c]{@{}c@{}}63→80,85,90\\ 63→95,100,105\end{tabular}} & 91.09\%                        & 91.37\%                & 90.02\%              \\ \hline
\end{tabular}
\begin{tablenotes}
\footnotesize
      \item [*] Two-Head B of MLMSA uses response and reliability SCI. Multi-Class B of SLMSA uses response and reliability SCI. 
      \item [**] ($m,cn$) means reliability SCI is obtained by repeating the measurements for $m$ times, which is divided into $cn$ categories/classes.
    \item [***] In the experiments of FF-APUFs, the responses of training set are obtained by a majority vote. While the responses of test set are noise-free.

\end{tablenotes}
\end{threeparttable}
}
\end{table*}

\section{Discussion}\label{sec:discussion}
\subsection{Different Loss Weights}

\begin{figure}[h]
	\centering
	\includegraphics[trim=0 0 0 0,clip,width=0.35\textwidth]{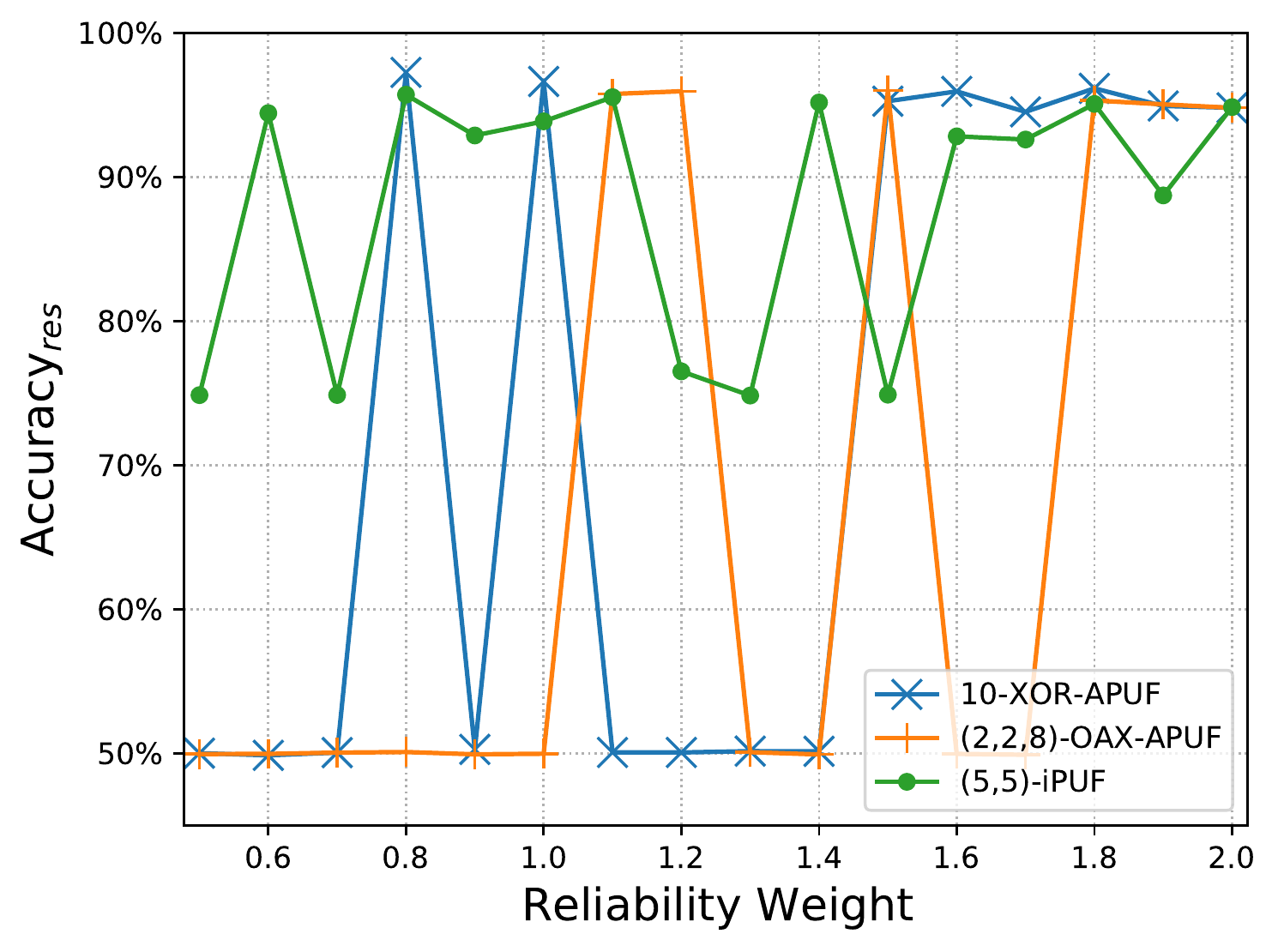}
	\caption{Response prediction accuracy under different loss settings. }
	\label{fig: lossweight results}
\end{figure}

We take Two-Head B of MLMSA attack to further explore the impact of different head loss weight settings on performance of the MLMSA. In this experiment, the response head loss weight is set to be $1$, while the reliability head loss weight ranges between $0.5$ and $2.0$. The attacked strong PUFs are $10$-XOR-APUF, ($2,2,8$)-OAX-APUF, and ($5,5$)-iPUF. As shown in Fig.~\ref{fig: lossweight results}, when the reliability head loss weight is small, the chance of response accuracy greater than 90\% tends to be small---each weight setting per strong PUF runs one time. The other observation is the attacking accuracy stability, for $10$-XOR-APUF and ($5,5$)-iPUF, when the reliability head loss weight is greater than $1.5$, the response prediction accuracy is high and stably maintained, e.g., above 90\% of $10$-XOR-APUF.

There are two general implications. Firstly, a slightly higher reliability head loss weight is necessary to enforce its contribution. Otherwise, if its weight is too small, the Two-Head B attack degrades to CRP-only based DL attacks, which is less effective exhibited by the lower chance of breaking large-scale APUF variants, e.g., 128-stage $10$-XOR-APUF. Secondly, a properly set higher head loss can make the attacking accuracy remain stably high with smaller variance, which can be observed by the accuracy of $10$-XOR-APUF and ($5,5$)-iPUF.

According to our observations on different loss curves during the training in all aforementioned experiments, the power output head converges the fastest, followed by the response output head, and finally the reliability output head. Though we always adopt a fixed head loss in all our experiments throughout the training process, it is expected that dynamically tuning these loss weights may achieve improved attack effect e.g., better accuracy or faster convergence for the total loss.

\subsection{Reliability Hybrid Attack}\label{sec:relattack}

The proposed MLMSA and reproduced SLMSA attack~\cite{liu2022multiclass} using the reliability SCI obtained from 11 repeated measurements can reliably model $10$-XOR-APUF with an accuracy of 96\%. As for ($2,2,8$)-OAX-APUF, the accuracy of the two attacks both reliably achieve 95\%. In fact, when these two attacks run for multiple times, there is a certain probability that the response accuracy of even larger scaled $11$, $12$-XOR-APUF, ($2,2,9$)-OAX-APUF can reach more than 90\% (i.e., being successfully broken). To be more precise, we have run 5 times, and there is one attempt reaching about 94\%.

As shown in Table~\ref{tab: attack results}, Two-Head B of MLMSA and Multi-Class B of SLMSA can successfully break $10$-XOR-APUF, ($2,2,8$)-OAX-APUF and ($5,5$)-iPUF when the reliability SCI assists the model training. In comparison, when only CRPs are used, the DL attack~\cite{santikellur2019deep} can only break $5$-XOR-APUF, ($2,2,3$)-OAX-APUF and ($4,4$)-iPUF, which are inferior to the hybrid attacks. Note to be fair, these attacks by us have used the same MLP structure except for the output layer. 

According to Liu \textit{et al.}~\cite{liu2022multiclass}, the GRA based hybrid attack using reliability SCI~\cite{tobisch2021combining} can successfully crack a 128-stage 6-XOR-APUF. But it cannot crack a $12$-XOR-APUF. GRA can break ($6,6$)-iPUF, which cannot be broken by the two attacks we have attempted. The GRA exhibits improved performance over iPUF due to its knowledge of the differential model for the iPUF. In addition, the GRA relies on a multiple pass attack when the $x>1$ in the ($x,y$)-iPUF.
In other words, the $y$-APUF are firstly learned, then the $x$-APUF are learned sequentially by fixing the learned $y$-APUF.

In summary, by using the easily obtainable reliability SCI to assist the hybrid modeling attacks, larger scaled strong APUF variants can be successfully broken, which \textit{cannot be achieved by using the CRP-only based DL attacks}.

 \begin{figure}
	\centering
	\subfigure[8-XOR-APUF]{
		\begin{minipage}[t]{0.33\linewidth}
			\centering
			\includegraphics[width=1.0\linewidth]{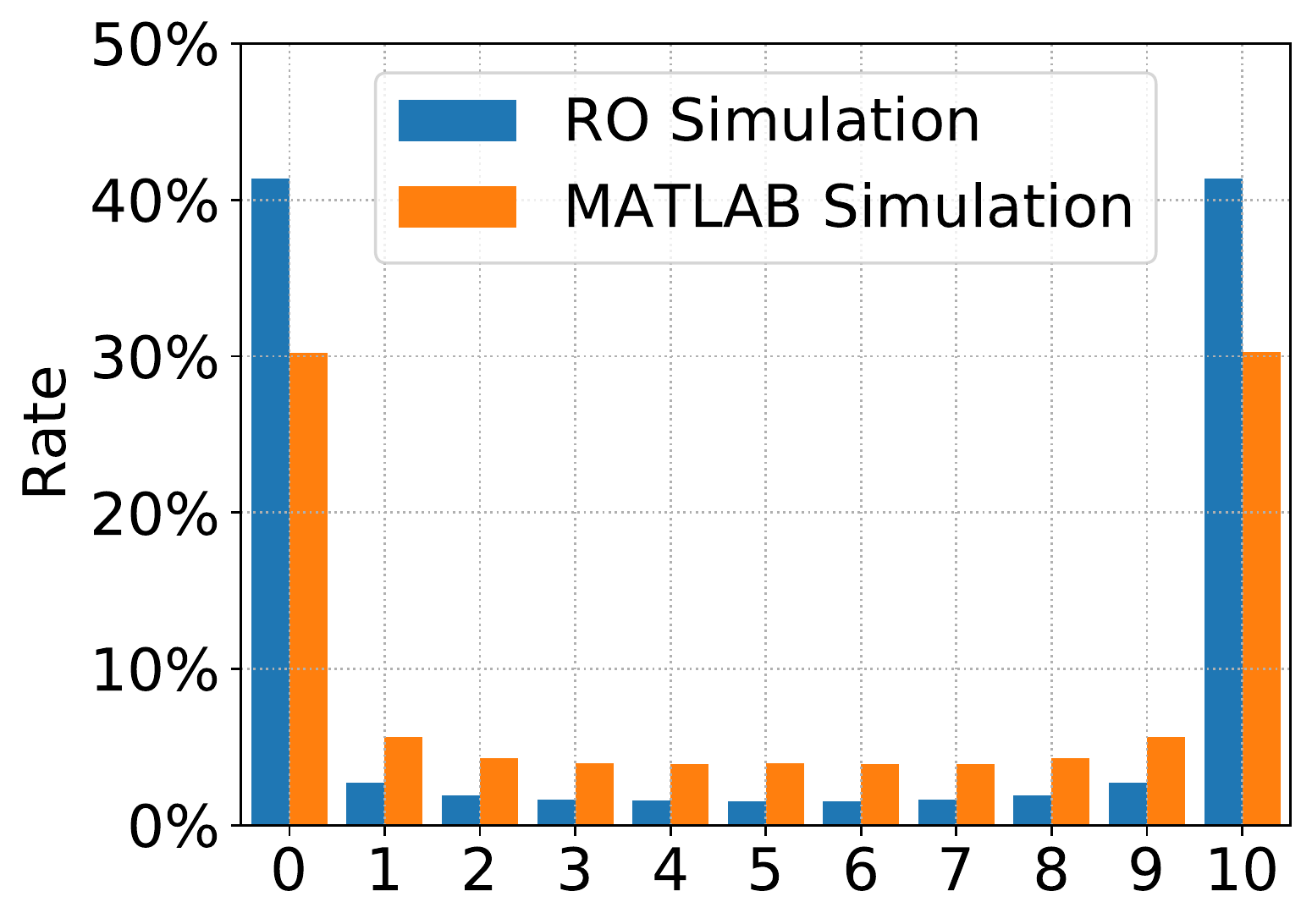}
		\end{minipage}
	}%
	\subfigure[9-XOR-APUF]{
		\begin{minipage}[t]{0.33\linewidth}
			\centering
			\includegraphics[width=1.0\linewidth]{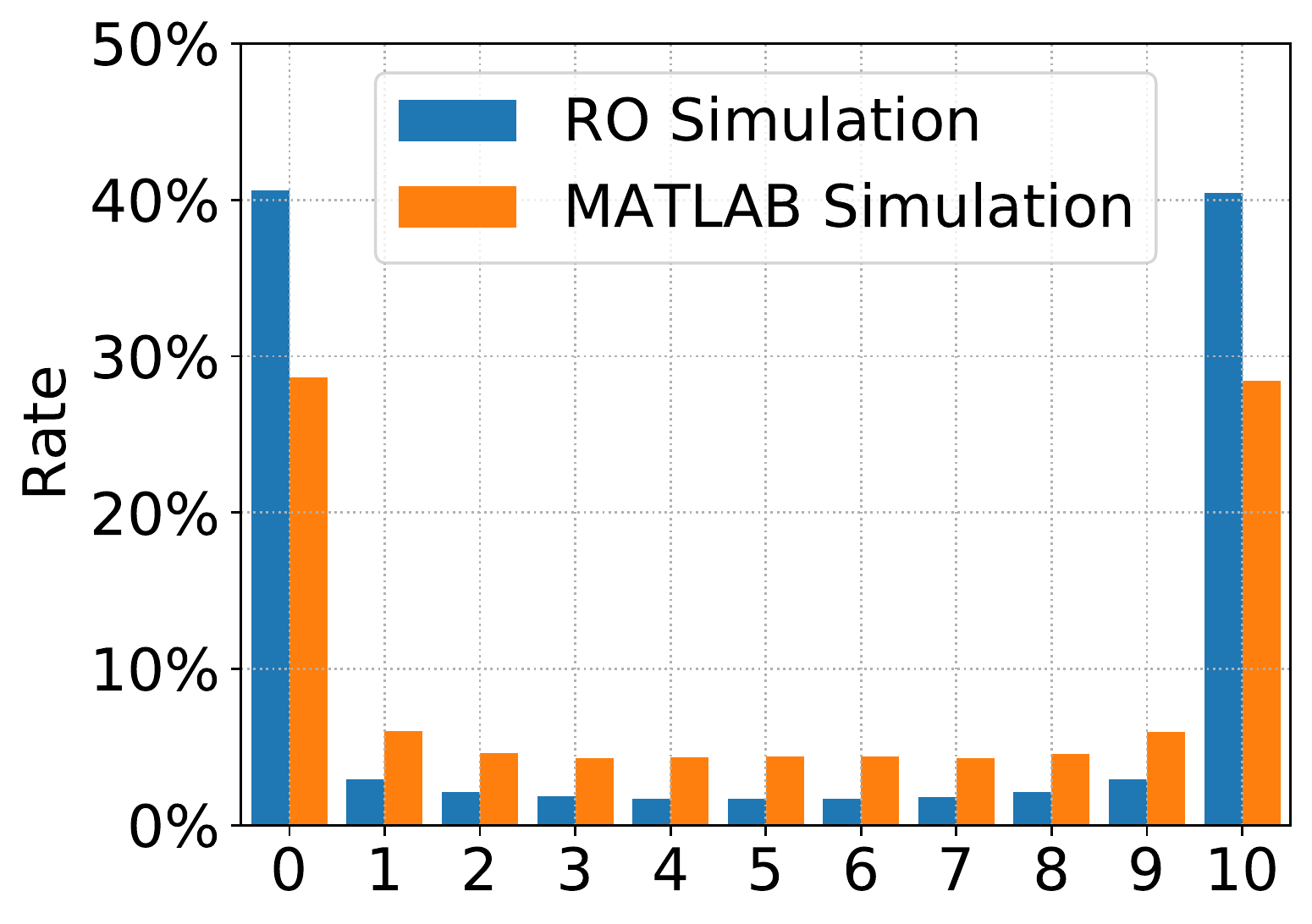}
		\end{minipage}
	}%
	\subfigure[10-XOR-APUF]{
		\begin{minipage}[t]{0.33\linewidth}
			\centering
			\includegraphics[width=1.0\linewidth]{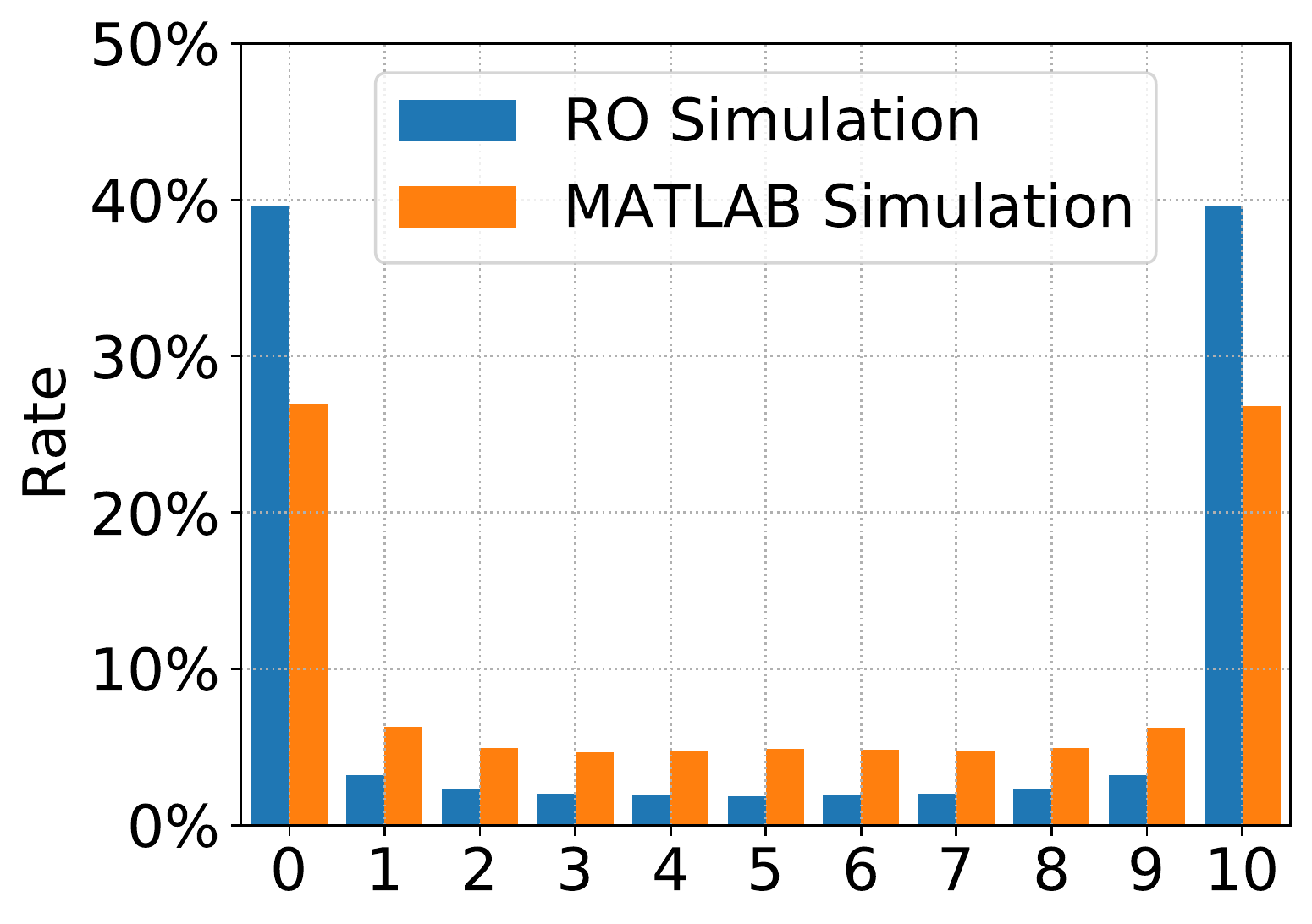}
		\end{minipage}
	}%
	
	\subfigure[(2,2,6)-OAX-APUF]{
		\begin{minipage}[t]{0.33\linewidth}
			\centering
			\includegraphics[width=1.0\linewidth]{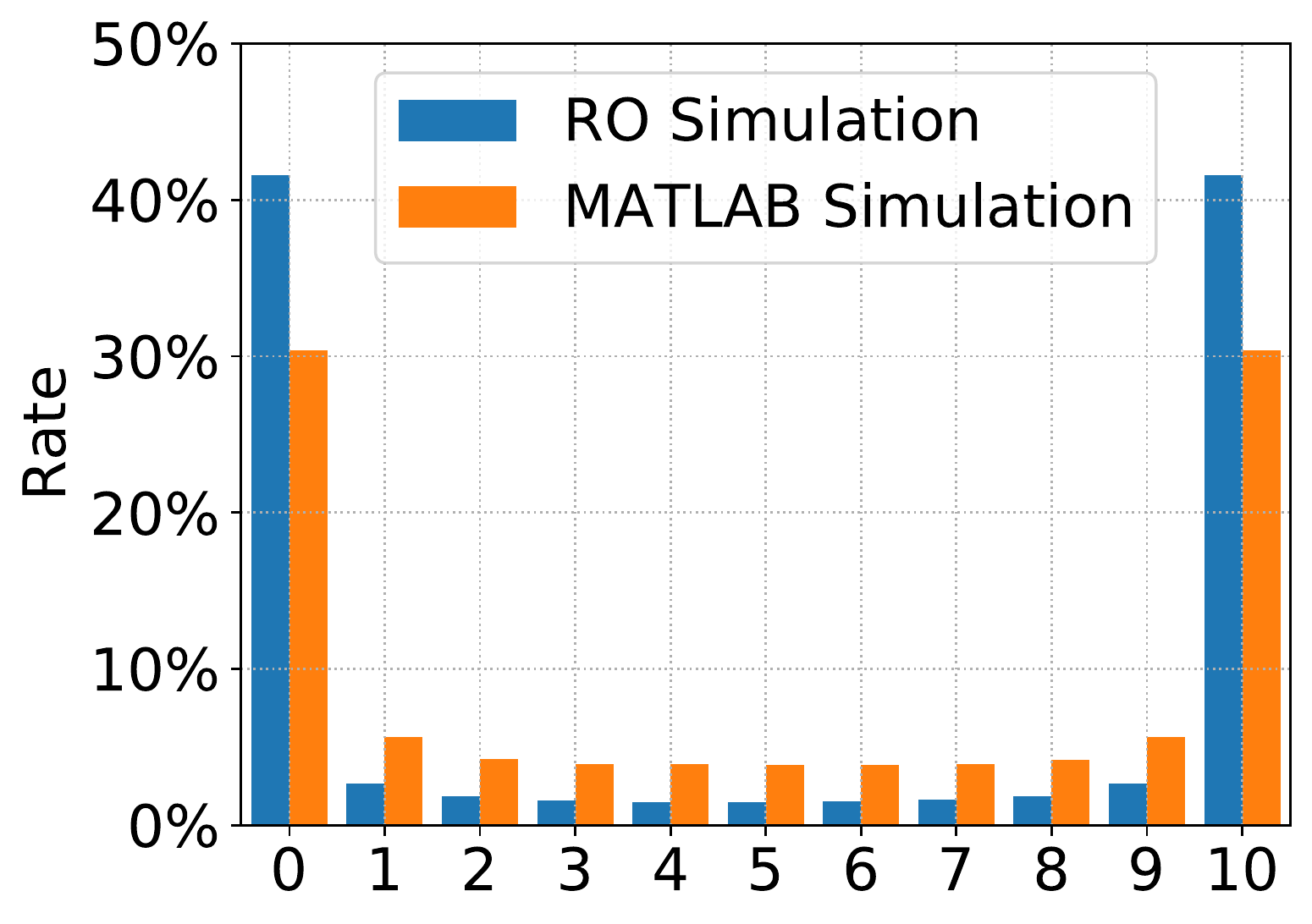}
		\end{minipage}
	}%
		\subfigure[(2,2,7)-OAX-APUF]{
		\begin{minipage}[t]{0.33\linewidth}
			\centering
			\includegraphics[width=1.0\linewidth]{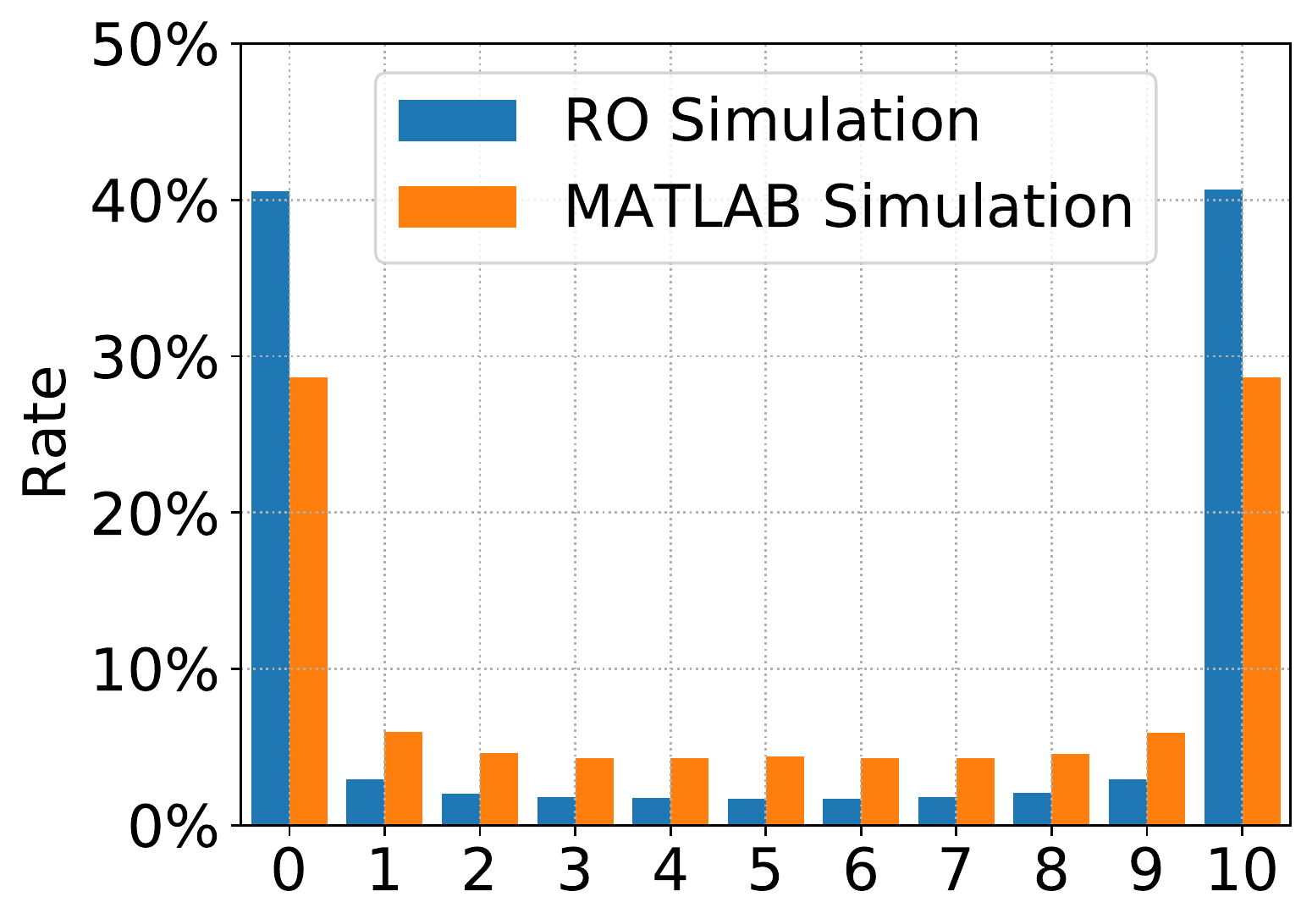}
		\end{minipage}
	}%
		\subfigure[(2,2,8)-OAX-APUF]{
		\begin{minipage}[t]{0.33\linewidth}
			\centering
			\includegraphics[width=1.0\linewidth]{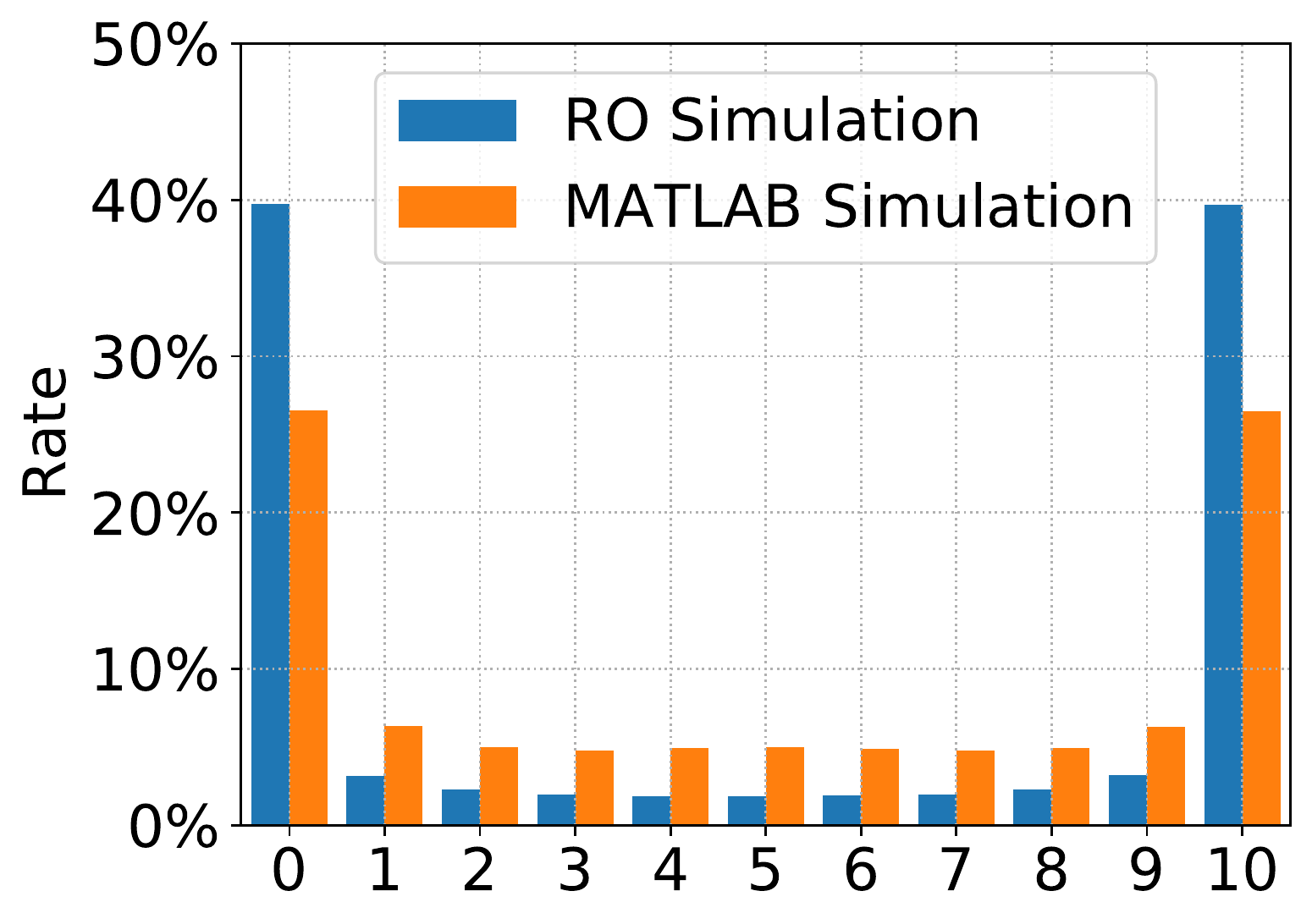}
		\end{minipage}
	}%
	\centering
	\caption{The response unreliability SCI category statistics of XOR-APUF (a-c) and OAX-APUF (d-f) based on silicon measurement sythesized RO-APUF and MATLAB numerical simulated APUF.}
	\label{fig:reliability statistics}
\end{figure}

\subsection{Silicon Measurement Validations}\label{sec:silicon}

\begin{figure}[h]
	\centering
	\includegraphics[trim=0 0 0 0,clip,width=0.25\textwidth]{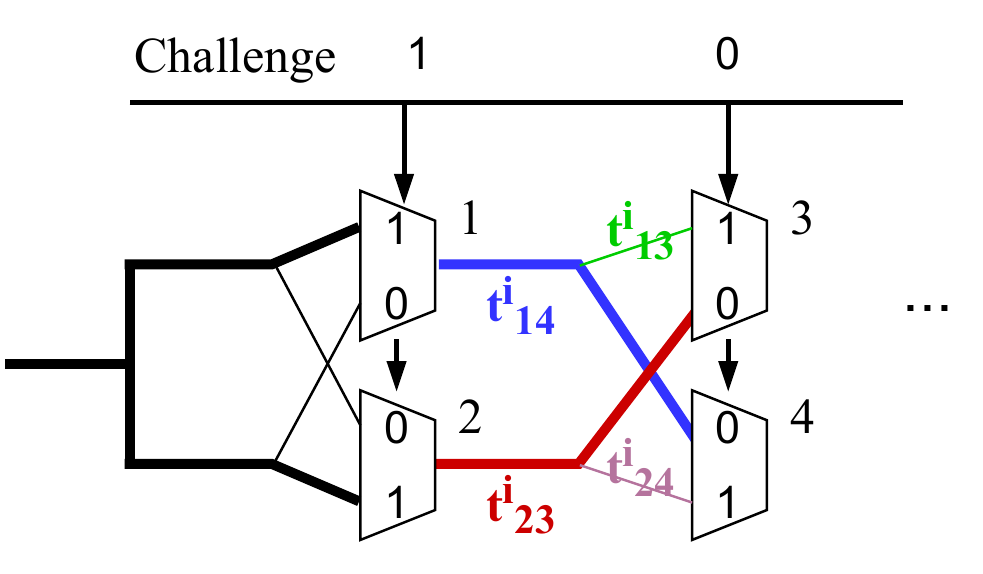}
	\caption{Example of $i$-th stage of an APUF with two signal paths (i.e., top and bottom paths). }
	\label{fig:roapuf stage}
\end{figure}

Following~\cite{gao2022treverse,gao2014highly}, we use the public ROPUF dataset HOST2018~\cite{hesselbarth2018large} to synthesize APUF, coined as RO-APUF. The key of this method is to use the reciprocal of four RO frequencies as the four time delays of each stage of the RO-APUF (see illustration in Fig.~\ref{fig:roapuf stage}). To synthesize a $128$-stage RO-APUF, $512$ RO frequencies are utilized, which mainly has the following three steps.

\begin{enumerate}
\item Obtain the reciprocal of RO frequencies to serve as the path segment delays of APUF: the reciprocal of four RO frequencies are used as the segment time delays of the $i$-th stage of APUF: t$_{13}^{i}$, t$_{14}^{i}$, t$_{23}^{i}$, t$_{24}^{i}$, as illustrated in Fig.~\ref{fig:roapuf stage}.

\item The $\rm delay\_cross^{i}=t_{14}^{i}-t_{23}^{i}$ and $\rm delay\_uncross^{i}=t_{13}^{i}-t_{24}^{i}$ are computed to represent the cross path delay difference and uncross path delay difference of the $i_{\rm th}$ stage. The $\boldsymbol{w}[i]$ is obtained through Eq.~\ref{wform}.

\begin{equation}
    \begin{split}
    \label{wform}
        &\boldsymbol{w}[0]=(\rm delay\_uncross^{0}-delay\_cross^{0})/2, \\
        &\boldsymbol{w}[128]=(\rm delay\_uncross^{127}+delay\_cross^{127})/2,\\
        &\boldsymbol{w}[i]=(\rm delay\_uncross^{i-1}+delay\_cross^{i-1}\\
        &+\rm delay\_uncross^{i}-delay\_cross^{i})/2,\\
        &i=1,2,...,127.
    \end{split}
\end{equation}

\item Compute the response of a given challenge according to Eq.~\ref{con:delta}, Eq.~\ref{con:feature vector}, and Eq.~\ref{con:responses}.
\end{enumerate}

The HOST2018 ROPUF dataset provides raw data from 217 Xilinx Artix-7 XC7A35T FPGAs, each containing a total of 6,592 ROs, comprising six different routing paths with 550 to 1,696 instances per type~\cite{hesselbarth2018large}. Each RO frequency is evaluated 100 times at 5\textcelsius, 15\textcelsius, 25\textcelsius, 35\textcelsius, 45\textcelsius, and 55\textcelsius. These repetitive frequency measurements are used for reliability SCI for the RO-APUF. However, this dataset does not have power consumption measurements---the silicon measured power SCI of the RO-APUF is unavailable. Therefore, the experiments below consider only the response and reliability SCI as inputs of the MLMSA when when compared with its SLMSA counterpart---noting SLMSA in~\cite{liu2022multiclass} is not evaluated against reliability SCI.

After synthesis, RO-APUF, XOR-RO-APUF and OAX-RO-APUF are used to validate the efficiency of the proposed MLMSA attack with silicon measurements. The reference response is measured at 25\textcelsius. The reliability SCI is measured $10$ times at 55\textcelsius. 

For the numerical simulated CRPs and the corresponding reliability SCI obtained through MATLAB simulator, the training size is $600,000$. Both MLMSA and SLMSA can successfully break $10$-XOR-APUF and ($2,2,8$)-OAX-APUF. However, the same training size and loss head weight settings are not directly applicable to XOR-RO-APUF and OAX-RO-APUF with silicon measurements. 

In order to achieve the same attack effect as the MATLAB numerical simulation, the training size and head loss weight settings are adjusted in some cases. When modeling the $l$-XOR-RO-APUF, the training size is adjusted to $600,000$ when $l\leq 7$; $1,200,000$ when $l=8,9$; and $1,500,000$ when $l=10$.
As for ($x=2,y=2,z$)-OAX-RO-APUF, the training size is adjusted to $600,000$ when $x+y+z \leq 9$; $1,200,000$ when $x+y+z=10,11$; $1,500,000$ when $x+y+z=12$. The response weight of all Two-Head B attacks on XOR-RO-APUF and OAX-RO-APUF are $1$.
For $l$-XOR-RO-APUF, the reliability loss weight is $0.8$ when $l=5,6,8,9$; $1.8$ when $l=7$; and $1$ when $l=10$. The reliability loss weight for the OAX-RO-APUF is always $0.8$. Both Two-Head B and Multi-Class B can model $10$-XOR-APUF with an accuracy of about 95\%. While Multi-Class B of SLMSA can not break ($2,2,8$)-OAX-RO-APUF, Two-Head B of MLMSA can successfully break it with an accuracy of 97.35\%.

Generally, the required number of training CRPs and corresponding reliability SCI silicon measurement attacks is larger than the number of numerical simulated strong PUFs.
The potential reason is that the unreliability of RO-APUF is lower than that from numerical simulation. The bit error rate or unreliability of RO-APUF is about 3\% to 5\%, while the unreliability of the APUF upon numerical simulation is about 5\% to 8\%. More precisely, as we repeatedly measure the same response $10$ times to gain $11$ categories for the reliability SCI, it indicates that the unreliable responses (categories of $0$ and $10$ are from those reliable responses, rest $1$ to $9$ categories are unreliable responses) in the RO-APUF is less than that in numerical simulation.
Therefore, the contribution from the reliability SCI is reduced, which requires a larger training size. Fig.~\ref{fig:reliability statistics} manifests this conjecture. We can see that the unreliable response categories of ($1$--$9$) of the silicon measurement based strong PUFs are much less than that from numerical simulations. 
Nonetheless, MLMSA and SLMSA can still successfully model XOR-RO-APUF and OAX-RO-APUF once the number of unreliable responses (i.e., those in categories $1$--$9$) increases.

\subsection{Curse of Dimensionality}\label{sec:curse}

\begin{figure}
  \centering
    \subfigure[Acc]{
     \label{fig:cell:a}
     \includegraphics[width=0.3\textwidth]{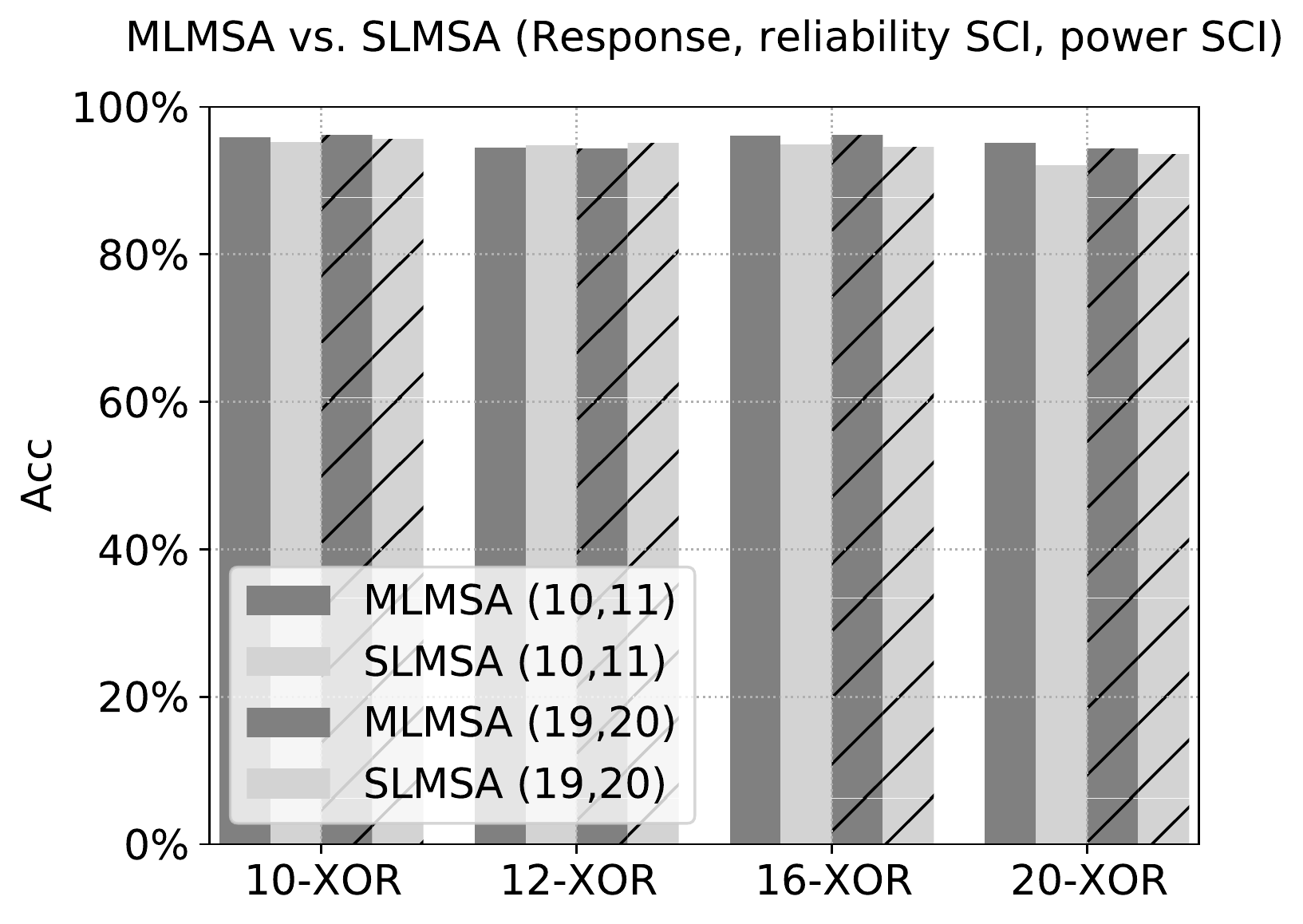}} 
    \subfigure[Time]{
     \label{fig:cell:b}
     \includegraphics[width=0.3\textwidth]{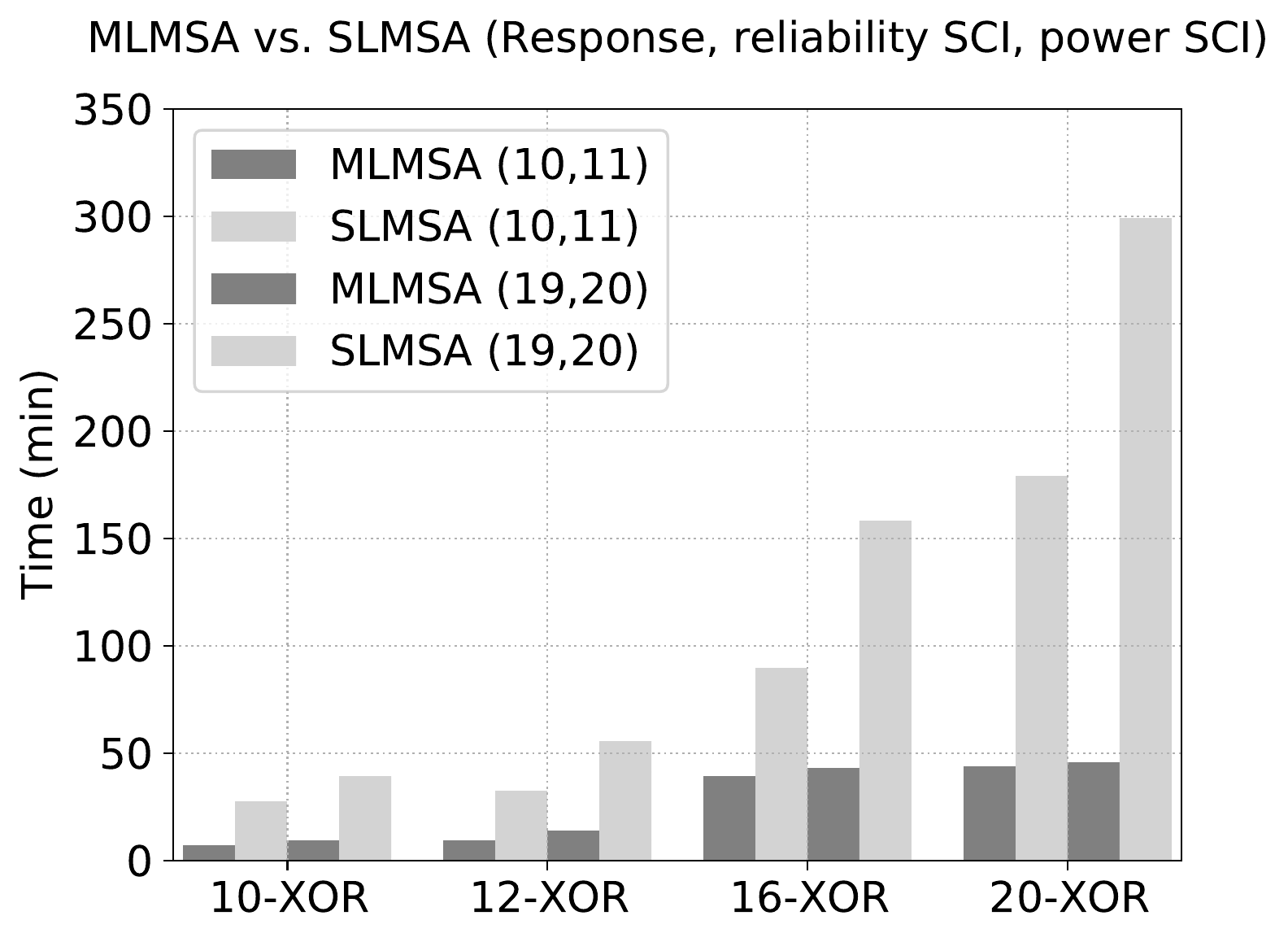}}
     \caption{MLMSA vs. SLMSA when the dimensionality is high (using response, reliability and power features).}
  \label{fig:cell}
 
\end{figure}

We now compare SLMSA with MLMSA when the large-scaled $l$-XOR-APUFs are attacked. In this context, when using all three response, reliability, and power SCI features, the dimensionality is high. The results are detailed in Table \ref{tab:two sci} and visualized in Fig.~\ref{fig:cell}.

As we can see, MLMSA always outperforms SLMSA when the dimensionality increases in terms of not only accuracy but also training time. The training time reduction is significant. When attacking $l$-XOR-APUF, the total dimensionality of SLMSA is $2\times cn \times (l+1)$ and MLMSA is $2+cn+(l+1)$. For instance, when attacking $20$-XOR-APUF using response, reliability SCI and power SCI, the dimensionality of SLMSA becomes 462 (i.e., $2\times 21 \times 11 = 462$) and 840 (i.e., $2\times 21 \times 20 = 840$) when ($m=10$, $cn=11$) and ($m=19$, $cn=20$), respectively. While for MLMSA, the summed dimensionality of its three output heads is $2+21+11=34$ and $2+21+20=44$, respectively, which is an order of magnitude smaller than that of SLMSA. The increased output dimensionality results in a larger network, which requires longer training time. In addition, when the number of classes of a single head greatly increases in SLMSA, the error increases---the classification hardness (slightly) goes up~\cite{fukunaga1984classification}.

\begin{table}[]
\caption{Comparison between MLMSA and SLMSA when attacking large-scaled $l$-XOR-APUFs resulting into high dimensionality by using response, reliability SCI and power SCI.}
\label{tab:two sci}
\centering
\resizebox{0.4\textwidth}{!}{
\begin{threeparttable}

\begin{tabular}{|c|c|c|c|c|c|}
\hline
\textbf{Attack}                 & \textbf{$l$} & \textbf{(m,cn)} & {\begin{tabular}[c]{@{}c@{}} Num. Classes \\ (Output Dimens.) \end{tabular}}& \textbf{Response Acc}          & \textbf{Time} \\ \hline
                                & 10              & (10,11)         & (2,11,11)            & 95.94\%                        & 7 min 2 s     \\ \cline{2-6} 
                                & 10              & (19,20)         & (2,11,20)            & 96.12\%                        & 9 min 36 s    \\ \cline{2-6} 
                                & 12              & (10,11)         & (2,13,11)            & 94.43\%                        & 9 min 33 s    \\ \cline{2-6} 
                                & 12              & (19,20)         & (2,13,20)            & 94.29\%                        & 13 min 39 s   \\ \cline{2-6} 
                                & 16              & (10,11)         & (2,17,11)            & 96.08\%                        & 39 min 21 s   \\ \cline{2-6} 
                                & 16              & (19,20)         & (2,17,20)            & 96.21\%                        & 42 min 51 s   \\ \cline{2-6} 
                                & 20              & (10,11)         & (2,21,11)            & 95.14\%                        & 43 min 49 s   \\ \cline{2-6} 
\multirow{-8}{*}{MLMSA}    & 20              & (19,20)         & (2,21,20)            & 94.28\%                        & 45 min 42 s   \\ \hline
                                & 10              & (10,11)         & 242                  & 95.26\% & 27 min 33 s   \\ \cline{2-6} 
                                & 10              & (19,20)         & 440                  & 95.66\% & 39 min 14 s   \\ \cline{2-6} 
                                & 12              & (10,11)         & 286                  & 94.77\% & 32 min 42 s   \\ \cline{2-6} 
                                & 12              & (19,20)         & 520                  & 95.08\% & 55 min 37 s   \\ \cline{2-6} 
                                & 16              & (10,11)         & 374                  & 94.92\% & 89 min 44 s   \\ \cline{2-6} 
                                & 16              & (19,20)         & 680                  & 94.55\% & 2 h 38 min    \\ \cline{2-6} 
                                & 20              & (10,11)         & 462                  & 92.12\%                        & 2 h 59 min    \\ \cline{2-6} 
\multirow{-8}{*}{SLMSA} & 20              & (19,20)         & 840                  & 93.58\%                        & 4 h 59 min    \\ \hline
\end{tabular}
\begin{tablenotes}
      \footnotesize
      \item [*] ($m,cn$) means reliability side-channel information is obtained by repeating the measurement for $m$ times, which is divided into $cn$ categories/classes.
      \item[**] SLMSA means the specific Multi-Class C that uses response, power SCI and reliability SCI to build CSPs.
      \item[***] MLMSA and SLMSA have the same number of hidden layers. The number of hidden layer is $3$ when $l=10,12,16$. The number of hidden layer is $4$ when $l=20$.
      \item[****] The Num.Classes (output dimensionality) for Three-head MLMSA means the number of classes of (response, power, reliability). The Num.Classes for SLMSA Multi-Class C means $2\times |power|\times |reliability|$.
\end{tablenotes}
\end{threeparttable}
}
\end{table}

\subsection{Lightweight Cryptographic Module Incorporation} The general take-away of this study is that by barely increasing the scale of a strong PUF, in particular, the APUF variants, is still challenging when it is confronted with rapidly evolving DL techniques, especially by combining response information and multiple SCIs. Therefore, the practical solution of using strong PUFs appears to incorporate lightweight cryptographic modules such as the Lockdown-PUF~\cite{yu2016lockdown}, TREVERSE constructions~\cite{gao2022treverse}, and RSO-APUF~\cite{zhang2020set} to protect the CRP interface. The overhead caused by lightweight cryptographic modules can be lower or comparable to the overhead incurred by increasing the APUF variants to a large-scale, e.g., 128-stage 30-XOR-APUF being breakable.

\subsection{MLP Insensitivity to Non-Uniformity}

Aghaie \textit{et al.}~\cite{aghaie2021inconsistency} demonstrated that implementation defects resulted in bias that hardens certain machine learning based modeling attacks (i.e., LR attacks) on complex PUF architectures (i.e., iPUF). Nonetheless, not any learning algorithm is sensitive to the APUF's non-uniformity when it is used to modeling the APUF variants. As recognized by Aghaie \textit{et al.}~\cite{aghaie2021inconsistency}, the ANN (in essence, the MLP is used) is insensitive to the non-uniformity based on their validations, which has been used to replace LR to mitigate the non-uniformity degradation on the attacking accuracy. In MLMSA, we have utilized a similar MLP learning algorithm.

\begin{figure}
	\centering
	\includegraphics[trim=0 0 0 0,clip,width=0.3\textwidth]{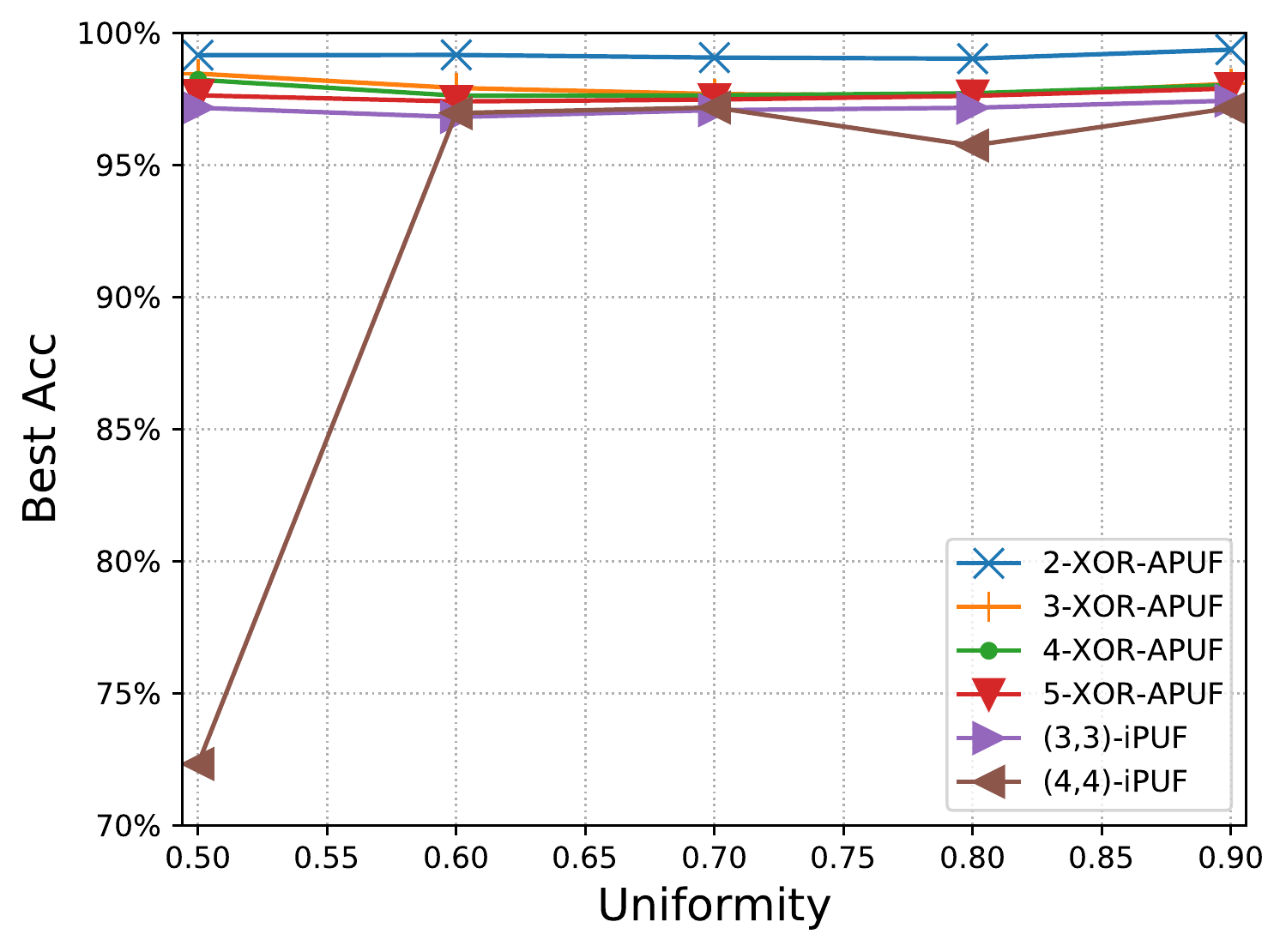}
	\caption{MLP attacking accuracy with biased $l$-XOR-APUFs and ($x,y$)-iPUFs.}
	\label{fig:bias}
\end{figure}

We have further evaluated the MLP's attacking accuracy when the underling PUFs' uniformity varies from 0.5 to 0.9 to check whether the MLP attack is indeed sensitive to the PUF uniformity.
The MLP structure, hyperparameter settings and training CRP size in the experiments follow~\cite{santikellur2019deep}. We note that the reported accuracy~\cite{santikellur2019deep} is hardly achieved until we replace the activation function \textsf{Relu} used in~\cite{santikellur2019deep} with \textsf{Tanh}. We note that the $l$-XOR-APUF with $l=3,4,5$ and $(3,3)$-iPUF, $(4,4)$-iPUF have been used by~\cite{aghaie2021inconsistency}, which we evaluate as well. As detailed in Fig.~\ref{fig:bias}, when the uniformity increases, the accuracy almost does not decline, but is even higher in some cases. Our results align with that of Aghaie \textit{et al.}~\cite{aghaie2021inconsistency}, that is the non-uniformity has negligible influence on increasing the APUF variants' modeling resilience to MLP. Therefore, MLMSA evaluations building upon the MLP model architecture hold their validity.

\begin{table}
\caption{Minimal training size of MLMSA to attack $l$-XOR-APUF---response, power, and reliability ($m=10$ and $cn=11$) features are used. The MLP has four hidden layers for $l=20$ and three hidden layers for the rest.}
\label{tab:xor m min}
\centering
\resizebox{0.4\textwidth}{!}{
\begin{tabular}{|c|c|c|c|c|c|c|}
\hline
$l$ & \begin{tabular}[c]{@{}c@{}}Training \\ Size\end{tabular} & \begin{tabular}[c]{@{}c@{}}Response \\ Weight\end{tabular} & \begin{tabular}[c]{@{}c@{}}Power \\ Weight\end{tabular} & \begin{tabular}[c]{@{}c@{}}Reliability \\ Weight\end{tabular} & Best Acc & \begin{tabular}[c]{@{}c@{}}Success\\  Rate\end{tabular} \\ \hline
5   & 50,000                                                   & 2                                                          & 10                                                      & 2                                                             & 92.91\%  & 10/10                                                   \\ \hline
6   & 62,000                                                   & 2                                                          & 10                                                      & 2                                                             & 93.24\%  & 10/10                                                   \\ \hline
7   & 68,000                                                   & 2                                                          & 10                                                      & 2                                                             & 91.82\%  & 8/10                                                    \\ \hline
8   & 80,000                                                   & 2                                                          & 10                                                      & 2                                                             & 91.94\%  & 8/10                                                    \\ \hline
9   & 92,000                                                   & 2                                                          & 10                                                      & 2                                                             & 91.66\%  & 6/10                                                    \\ \hline
10  & 100,000                                                  & 2                                                          & 10                                                      & 2                                                             & 92.12\%  & 7/10                                                    \\ \hline
11  & 145,000                                                  & 2                                                          & 10                                                      & 2                                                             & 94.01\%  & 10/10                                                   \\ \hline
12  & 164,000                                                  & 4                                                          & 5                                                       & 2                                                             & 93.98\%  & 10/10                                                   \\ \hline
13  & 164,000                                                  & 2                                                          & 10                                                      & 2                                                             & 93.52\%  & 6/10                                                    \\ \hline
14  & 200,000                                                  & 4                                                          & 5                                                       & 2                                                             & 93.72\%  & 9/10                                                    \\ \hline
15  & 240,000                                                  & 4                                                          & 5                                                       & 2                                                             & 93.10\%  & 8/10                                                    \\ \hline
16  & 260,000                                                  & 10                                                         & 8                                                       & 2                                                             & 93.09\%  & 7/10                                                    \\ \hline
17  & 280,000                                                  & 5                                                          & 5                                                       & 2                                                             & 93.41\%  & 6/10                                                    \\ \hline
18  & 300,000                                                  & 5                                                          & 5                                                       & 2                                                             & 93.51\%  & 9/10                                                    \\ \hline
19  & 340,000                                                  & 10                                                         & 8                                                       & 2                                                             & 93.18\%  & 7/10                                                    \\ \hline
20  & 360,000                                                  & 10                                                         & 5                                                       & 2                                                             & 93.88\%  & 9/10                                                    \\ \hline
\end{tabular}

}
\end{table}

\begin{table}
\caption{Minimal training size of MLMSA to attack ($x,y$)-iPUF---response, power, and reliability ($m=10$ and $cn$=11) features are used. The MLP has four hidden layers.}
\label{tab:ipuf m min}
\centering
\resizebox{0.4\textwidth}{!}{
\begin{tabular}{|c|c|c|c|c|c|c|}
\hline
(x,y)  & \begin{tabular}[c]{@{}c@{}}Training \\ Size\end{tabular} & \begin{tabular}[c]{@{}c@{}}Response\\  Weight\end{tabular} & \begin{tabular}[c]{@{}c@{}}Reliability\\  Weight\end{tabular} & \begin{tabular}[c]{@{}c@{}}Power \\ Weight\end{tabular} & Best Acc & \begin{tabular}[c]{@{}c@{}}Success \\ Rate\end{tabular} \\ \hline
(4,4)  & 100,000                                                  & 2                                                          & 2                                                             & 10                                                      & 93.66\%  & 9/10                                                    \\ \hline
(5,5)  & 130,000                                                  & 2                                                          & 2                                                             & 10                                                      & 94.35\%  & 10/10                                                   \\ \hline
(1,10) & 160,000                                                  & 2                                                          & 2                                                             & 10                                                      & 91.61\%  & 7/10                                                    \\ \hline
(6,6)  & 170,000                                                  & 2                                                          & 2                                                             & 10                                                      & 94.17\%  & 6/10                                                    \\ \hline
(7,7)  & 200,000                                                  & 2                                                          & 2                                                             & 10                                                      & 93.61\%  & 8/10                                                    \\ \hline
(8,8)  & 250,000                                                  & 2                                                          & 2                                                             & 10                                                      & 93.55\%  & 10/10                                                   \\ \hline
(2,16) & 480,000                                                  & 2                                                          & 2                                                             & 10                                                      & 91.19\%  & 6/10                                                    \\ \hline
\end{tabular}

}
\end{table}

\begin{table}
\caption{Minimal training size generally decreases given higher trainable parameters of MLP in the MLMSA attack against $l$-XOR-APUF and ($x,y$)-iPUF.}
    \label{tab: diffHiddenLayer}
    \centering
    \resizebox{0.4\textwidth}{!}{
\begin{tabular}{|c|c|c|c|c|c|c|c|}
\hline
PUF        & \begin{tabular}[c]{@{}c@{}}Training \\ Size\end{tabular} & \begin{tabular}[c]{@{}c@{}}Hidden \\ Layer\\ Num\end{tabular} & \begin{tabular}[c]{@{}c@{}}Response \\ Weight\end{tabular} & \begin{tabular}[c]{@{}c@{}}Power \\ Weight\end{tabular} & \begin{tabular}[c]{@{}c@{}}Reliability \\ Weight\end{tabular} & Best Acc & \begin{tabular}[c]{@{}c@{}}Success\\  Rate\end{tabular} \\ \hline
5-XOR-APUF & 80,000                                                   & 2                                                           & 2                                                          & 10                                                      & 2                                                             & 92.15\%  & 8/10                                                    \\ \hline
5-XOR-APUF & 50,000                                                   & 3                                                           & 2                                                          & 10                                                      & 2                                                             & 92.91\%  & 10/10                                                   \\ \hline
5-XOR-APUF & 45,000                                                   & 4                                                           & 2                                                          & 10                                                      & 2                                                             & 92.99\%  & 9/10                                                    \\ \hline
(4,4)-iPUF & 110,000                                                  & 3                                                           & 2                                                          & 10                                                      & 2                                                             & 94.49\%  & 10/10                                                   \\ \hline
(4,4)-iPUF & 100,000                                                  & 4                                                           & 2                                                          & 10                                                      & 2                                                             & 93.66\%  & 9/10                                                    \\ \hline
(4,4)-iPUF & 100,000                                                  & 5                                                           & 2                                                          & 10                                                      & 2                                                             & 94.28\%  & 9/10                                                    \\ \hline
(4,4)-iPUF & 90,000                                                   & 5                                                           & 2                                                          & 10                                                      & 2                                                             & 93.85\%  & 7/10                                                    \\ \hline
\end{tabular}
}
    
\end{table}

\begin{table}
\caption{Increase reliability granularity helps to improve the MLMSA attack accuracy given the same training size. The $l$-XOR-RO-APUF is attacked with a $600,000$ training size (response and reliability feature are used). The MLP has three hidden layers.}
\label{tab:xor m mlmsa}
\centering
\resizebox{0.4\textwidth}{!}{
\begin{tabular}{|cccccc|}
\hline
\multicolumn{1}{|c|}{m}  & \multicolumn{1}{c|}{cn} & \multicolumn{1}{c|}{5-XOR-APUF}    & \multicolumn{1}{c|}{6-XOR-APUF}    & \multicolumn{1}{c|}{7-XOR-APUF}    & 8-XOR-APUF    \\ \hline
\multicolumn{1}{|c|}{10} & \multicolumn{1}{c|}{11} & \multicolumn{1}{c|}{98.38\%} & \multicolumn{1}{c|}{98.13\%} & \multicolumn{1}{c|}{50.25\%} & 49.95\% \\ \hline
\multicolumn{1}{|c|}{20} & \multicolumn{1}{c|}{21} & \multicolumn{1}{c|}{98.30\%} & \multicolumn{1}{c|}{98.39\%} & \multicolumn{1}{c|}{97.72\%} & 50.02\% \\ \hline
\multicolumn{1}{|c|}{30} & \multicolumn{1}{c|}{31} & \multicolumn{1}{c|}{98.52\%} & \multicolumn{1}{c|}{97.93\%} & \multicolumn{1}{c|}{97.75\%} & 50.10\% \\ \hline
\multicolumn{1}{|c|}{40} & \multicolumn{1}{c|}{41} & \multicolumn{1}{c|}{98.37\%} & \multicolumn{1}{c|}{98.22\%} & \multicolumn{1}{c|}{97.99\%} & 50.08\% \\ \hline
\multicolumn{1}{|c|}{50} & \multicolumn{1}{c|}{51} & \multicolumn{1}{c|}{98.35\%} & \multicolumn{1}{c|}{97.97\%} & \multicolumn{1}{c|}{97.98\%} & 97.65\% \\ \hline
\end{tabular}
}
\end{table}

\begin{table}
\caption{Increase reliability granularity helps to improve the MLMSA attack accuracy given the same training size. The ($x,y,z$)-OAX-RO-APUF is attacked with a $600,000$ training size (response and reliability feature are used).}
\label{tab:oax m mlmsa}
\centering
\resizebox{0.4\textwidth}{!}{
\begin{tabular}{|ccccccc|}
\hline
\multicolumn{1}{|c|}{m}  & \multicolumn{1}{c|}{cn} & \multicolumn{1}{c|}{(2,2,2)} & \multicolumn{1}{c|}{(2,2,3)} & \multicolumn{1}{c|}{(2,2,4)} & \multicolumn{1}{c|}{(2,2,5)} & (2,2,6) \\ \hline
\multicolumn{1}{|c|}{10} & \multicolumn{1}{c|}{11} & \multicolumn{1}{c|}{98.38\%} & \multicolumn{1}{c|}{98.07\%} & \multicolumn{1}{c|}{97.56\%} & \multicolumn{1}{c|}{97.19\%} & 49.92\% \\ \hline
\multicolumn{1}{|c|}{20} & \multicolumn{1}{c|}{21} & \multicolumn{1}{c|}{98.40\%} & \multicolumn{1}{c|}{98.22\%} & \multicolumn{1}{c|}{97.75\%} & \multicolumn{1}{c|}{97.46\%} & 50.31\% \\ \hline
\multicolumn{1}{|c|}{30} & \multicolumn{1}{c|}{31} & \multicolumn{1}{c|}{98.50\%} & \multicolumn{1}{c|}{98.19\%} & \multicolumn{1}{c|}{97.82\%} & \multicolumn{1}{c|}{97.64\%} & 50.33\% \\ \hline
\multicolumn{1}{|c|}{40} & \multicolumn{1}{c|}{41} & \multicolumn{1}{c|}{98.50\%} & \multicolumn{1}{c|}{98.37\%} & \multicolumn{1}{c|}{97.89\%} & \multicolumn{1}{c|}{97.47\%} & 51.46\% \\ \hline
\multicolumn{1}{|c|}{50} & \multicolumn{1}{c|}{51} & \multicolumn{1}{c|}{98.33\%} & \multicolumn{1}{c|}{98.33\%} & \multicolumn{1}{c|}{97.89\%} & \multicolumn{1}{c|}{97.42\%} & 97.22\% \\ \hline
\end{tabular}
}
\end{table}

\begin{figure}
	\centering
	\includegraphics[trim=0 0 0 0,clip,width=0.30\textwidth]{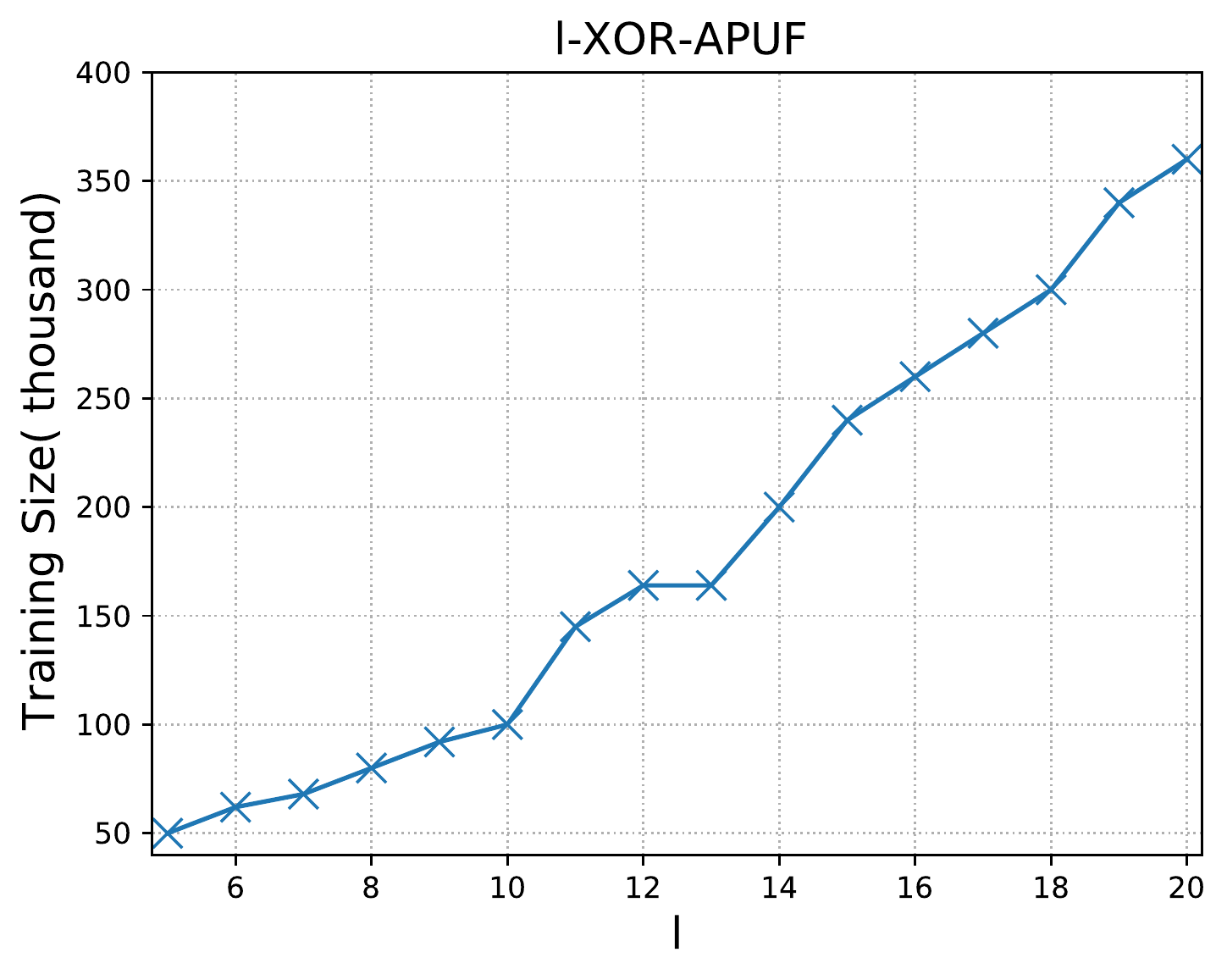}
	\caption{The minimal training size of MLMSA attack on $l$-XOR-APUFs.}
	\label{fig:xor min}
\end{figure}

\subsection{Minimal Training Size}

It is preferable to find the minimal number of training CRPs to comprehend the robustness of PUFs from a theoretical (i.e., PAC Learning~\cite{ganji2016pac}) or empirical (i.e., Intrinsic ID~\cite{santikellur2021correlation}) point of view to demonstrate the attack efficiency from another perspective. For machine learning based APUF modeling attacks, it is non-trivial to gain a theoretical bound of the minimal training size. This is especially the case for MLMSA that leverages multiple features given the same challenge. We have aimed at empirically finding the minimal training size with the same MLP trainable size and reliability granularity. The results are shown in Table~\ref{tab:xor m min} and Table~\ref{tab:ipuf m min} against $l$-XOR-APUF and ($x,y$)-iPUF, respectively. Note that when the PUF scale goes up, the minimal training size does gradually increase, see the visualization in Figure~\ref{fig:xor min} with a much small slope.
This minimal training size increase is slight (i.e., very small slope) compared to that reported by Becker~\cite{becker2015gap} that used reliability-based attack incorporating a divide-and-conquer strategy to break underlying APUFs within $l$-XOR-APUF one-by-one---the CMA-ES algorithm is leveraged. In this case, the attacking complexity (i.e., attacking time and required number of CRPs) is also (approximately) linear with $l$. We note that the CMA-ES will find each APUF's numerical weights that have a linear relationship with the real time-delay segments of the APUF. In other words, it needs to recover the mathematical model per APUF.

We further note that the implemented MLP model's performance is dependent on its trainable parameters. That is, by properly tuning the MLP trainable size, the performance (i.e., the attacking accuracy) can be improved even given the same training size---see experimental results in Table~\ref{tab: diffHiddenLayer}. For example, when the $5$-XOR-APUF is attacked, the minimal training size is reduced from 80,000 to 45,000 when more hidden layers are leveraged in the MLP---higher number of trainable parameters.

Moreover, when the reliability SCI is used, the minimal training size can be reduced if the granularity (i.e., $m$, and $cn$) of the reliability increases---see experimental results in Table~\ref{tab:xor m mlmsa} for attacking $l$-XOR-APUF and Table~\ref{tab:oax m mlmsa} for attacking ($x,y,z$)-OAX-APUF. For example, in Table~\ref{tab:xor m mlmsa}, when the reliability granularity is low ($m=10$ and $cn=11$), the $7$-XOR-APUF and $8$-XOR-APUF cannot be broken. When $m=20$ and $cn=21$, the $7$-XOR-APUF is breakable with the same training size but increased reliability granularity. When $m=50$ and $cn=51$, the $8$-XOR-APUF is further broken with the same training size but a further increased reliability granularity.

\subsection{Limitations}

It has been found that the modeling attack results of simulated CRPs can (reasonably) reflect the APUF variants modeling resilience based on silicon CRPs~\cite{ruhrmair2013puf}. Using simulated CRPs following the well-established linear additive delay model~\cite{lim2005extracting,ruhrmair2013puf,delvaux2013side,becker2015pitfalls,shi2019approximation,sahoo2017multiplexer,wisiol2019breaking,wisiol2022neural} to evaluate the \textit{modeling resilience} is a common and acceptable means, despite other parameters (i.e., uniformity and uniqueness) sometimes exhibiting some inconsistency when the physical implementation is not properly tuned (i.e., some paths are extremely asymmetric, resulting in severe bias).

Silicon measurement based validations are preferable and have been conducted in other modeling resilience studies~\cite{ruhrmair2013puf,ruhrmair2014efficient,nguyen2019interpose}. It is worth validating the MLMSA and SLMSA with silicon measurements, especially, for the power SCI in future work. The absence of silicon fabrication based validation is a limitation of current study.

To remedy this limitation to some extent, following~\cite{gao2014highly,gao2022treverse}, we have evaluated the MLMSA efficacy through silicon measurement via the synthesized RO-APUFs---the response and reliability SCI are emulated in this context. The characteristics are the same for the numerical simulation based evaluations.

The weights or regularization factors of response and other SCIs (i.e., power and reliability) in the MLMSA are empirically determined, which have shown sufficient attacking performance. It is interesting to incorporate adaptive weights learning during the training to automate their optimization in future work.

\section{Conclusion} \label{sec:conclusion}
This work proposes the MLMSA attack that constructively leverages multi-head DL to concurrently exploit useful multi-channel information to attack strong PUFs, particularly, APUF variants. With this simple and efficient MLMSA attack, we have successfully attacked
a 128-stage 30-XOR-APUF, a (9, 9)- and (2, 18)-iPUF, and a (2, 2, 30)-OAX-APUF when CRPs, power SCI and reliability SCI are simutaneously used. With access to only easy-to-obtain reliability SCI and CRPs, the MLMSA can stably break a 128-stage $10$-XOR-APUF, ($2,2,8$)-OAX-APUF, and $6$-loop FF-APUF; and statistically break a $12$-XOR-APUF and ($2,2,9$)-OAX-APUF. All these large-scaled strong APUF variants have not been achieved by state-of-the-arts attacks. We conclude that MLMSA can serve as an efficient technique for examining other existing or emerging strong PUF's modeling resilience due to its simplicity, efficacy and the avoidance of underlying mathematical model. 


\bibliographystyle{IEEEtran}
\bibliography{References}

\end{document}